A method for quantifying sectoral optic disc pallor in fundus photographs and its association with peripapillary RNFL thickness

Samuel Gibbon,[1,2] Graciela Muniz-Terrera,[3] Fabian SL Yii,[1,2] Charlene Hamid,[1] Simon Cox,[5] Ian JC Maccormick,[6,7] Andrew J Tatham,[1,8] Craig Ritchie,[1,3] Emanuele Trucco,[9] Baljean Dhillon[1,8] Thomas J MacGillivray[1,2,4]

[1]Centre for Clinical Brain Sciences, Edinburgh, UK

[2]Robert O Curle Ophthalmology Suite, Institute for Regeneration and Repair, University of Edinburgh, UK

[3]Centre for Dementia Prevention, University of Edinburgh, Edinburgh, UK

[4]VAMPIRE project, Edinburgh Clinical Research facility, University of Edinburgh

[5]Lothian Birth Cohorts, Department of Psychology, University of Edinburgh, Edinburgh, UK

[6]Centre for Inflammation Research, UoE, UK

[7]Institute for Adaptive and Neural Computation, UoE, UK

[8]Princess Alexandra Eye Pavilion, Chalmers Street, Edinburgh, UK

[9]VAMPIRE project, Computing (SSEN), University of Dundee, Dundee, UK

**Abstract**

*Purpose:* To develop an automatic method of quantifying optic disc pallor in fundus photographs and determine associations with peripapillary retinal nerve fibre layer (pRNFL) thickness.

*Methods:* We used deep learning to segment the optic disc, fovea, and vessels in fundus photographs, and measured pallor. We assessed the relationship between pallor and pRNFL thickness derived from optical coherence tomography scans in 118 participants. Separately, we used images diagnosed by clinical inspection as pale (N=45) and assessed how measurements compared to healthy controls (N=46). We also developed automatic rejection thresholds, and tested the software for robustness to camera type, image format, and resolution.

*Results:* We developed software that automatically quantified disc pallor across several zones in fundus photographs. Pallor was associated with pRNFL thickness globally ($\beta$ = -9.81 (SE = 3.16), $p < 0.05$), in the temporal inferior zone ($\beta$ = -29.78 (SE = 8.32), $p < 0.01$), with the nasal/temporal ratio ($\beta$ = 0.88 (SE = 0.34), $p < 0.05$), and in the whole disc ($\beta$ = -8.22 (SE = 2.92), $p < 0.05$). Furthermore, pallor was significantly higher in the patient group. Lastly, we demonstrate the analysis to be robust to camera type, image format, and resolution.

*Conclusions:* We developed software that automatically locates and quantifies disc pallor in fundus photographs and found associations between pallor measurements and pRNFL thickness.

*Translational relevance:* We think our method will be useful for the identification, monitoring and progression of diseases characterized by disc pallor/optic atrophy, including glaucoma, compression, and potentially in neurodegenerative disorders.

**Introduction**

A pale optic disc is the hallmark of optic atrophy, which refers to the irreversible loss or damage of retinal ganglion cell (RGC) axons along the anterior visual pathway.[1] A pale disc has numerous potential causes, including inflammation, ischemia, compression, raised intracranial pressure, toxicity, nutritional deficiency, trauma, hereditary conditions, vascular disease, infection, and retinal disease.[1,2] As such, disc pallor indicates the end stage of one of several disease processes. It begins to show around 4-6 weeks after axonal damage.[1] In clinical practice, a pale disc is often considered to be due to a compressive lesion until further tests prove otherwise.[1,3] Correctly identifying disc pallor can lead to life-saving treatment.

Pallor can be identified through ophthalmoscopy or fundus photography. However, these methods are limited in that assessing change over time may be difficult, judgement can vary substantially among observers,[4] and the location of pallor is often not consistently recorded.[5] Computational approaches may offer a solution, but efforts have been limited, either requiring special filters during acquisition,[6–8] or manual demarcation of the disc.[9,10] The purpose of this study was to develop a fully automatic method of locating and quantifying disc pallor in fundus photographs.

Damage to RGC axons can be observed directly with optical coherence tomography (OCT), which images the retinal nerve fibre layer (RNFL). Accordingly, we validated the tool by comparing pallor with peripapillary (pRNFL) thickness from OCT scans in anatomically equivalent zones. Additionally, we tested the software on an image set with clinically diagnosed pallor. Finally, we tested the robustness of the software to camera type, image format, and resolution with a variety of data sets.

## 2. Materials and methods

*Participants and image capture*

Several datasets were used in model development and testing (Figure 1), however, only one dataset (PREVENT) was used for testing the association between pallor and pRNFL thickness. The PREVENT-Dementia study protocol is described elsewhere.[11] Briefly, participants aged 40 to 59 were recruited through multiple sources from five sites in the UK. Retinal imaging was conducted in a sub-study at the Edinburgh site only (N=123), which included fundus photography centred halfway between the optic disc and the macula (Figure 2, right), with a non-mydriatic 45-degree field of view camera (CR-DGi; Canon USA, Inc., Lake Success, NY), and OCT (Heidelberg SPECTRALIS, Heidelberg Engineering, Germany; Figure 2, 4 leftmost images). The N-site circular scan OCT module was used, set to high speed (1,536 A-scans) with a target Automatic Real Time-function (ART) of 100. Participants provided written informed consent, and the study was carried out in compliance with the Declaration of Helsinki. Six additional datasets were used in model development and testing (Table 1).

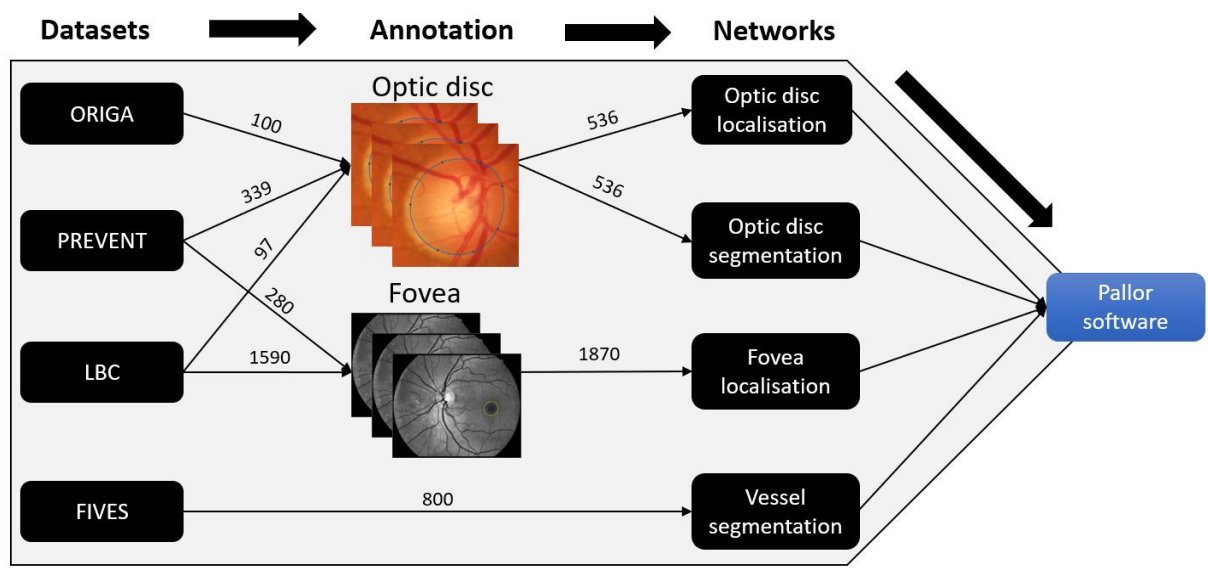

**Figure 1**. Flow of images through annotation, networks, and into the final software.

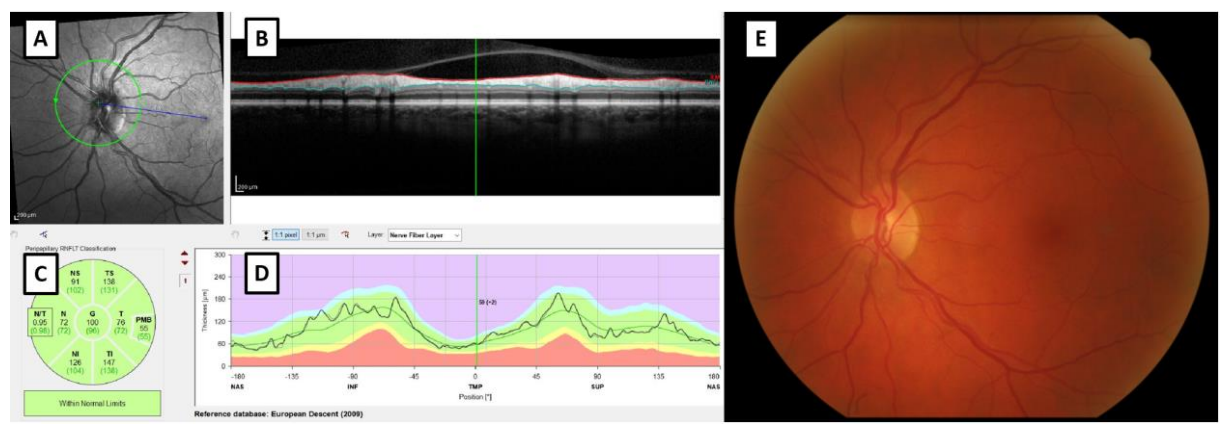

**Figure 2**. A, B, C and D represent output of the Heidelberg SPECTRALIS peripapillary scan, where A shows the scan location, B shows the various layers (thick topmost layer is RNFL), C shows the measurement zones, and D shows the normative data (jagged line is the current participant). E shows the corresponding retinal fundus image.

**Table 1**. Characteristics of datasets used in model development and testing.

| Variable | PREVENT[12] | LBC 1936[13] | ORIGA[14] | FIVES[15] | RIM-ONE[16] | IDRiD[17] | RFMiD[18] |
|---|---|---|---|---|---|---|---|
| Number of images | 339* | 1623 | 650 | 800 | 169 (v1) 485 (v3) | 103 | 92 |
| Number of images rejected | 2 | 33 | 0 | 0 | 0 | 0 | 1 |
| Camera | Canon CR-Dgi | Canon CR-Dgi | Canon CR-Dgi | Topcon TRC-NW8 | Nidek AFC-210 & Kowa WX sD | Kowa VX-10α | TOPCON sD OCT-2000, Kowa VX-10α, TOPCON TRC-NW300 |
| FOV | 45 | 45 | -- | 50 | Nidek = 45, Kowa = 34 | 50 | 45 – 50 |
| Resolution | 3072 x 2048 | 2689 x 2186 | 2518 x 2048 | 2048 x 2048 | Between 318 x 318 and 626 x 626 | 4288 x 2848 | Between 2144 x 1424 and 4288 x 2848 |
| Format | BMP | BMP | JPEG | PNG | PNG | JPG | PNG |
| Country | UK | UK | Singapore | China | Spain | India | India |
| Age (mean, SD) / range | 51.5 (5.5) | 72.5 (0.7) | 40-79 | 4-83 | -- | -- | -- |
| Sex (F) | 57.70% | 49.80% | -- | -- | -- | -- | -- |
| Purpose of data collection | Cohort study in dementia | Population-based study of cognitive ageing | Population-based study of eye diseases | Enable AI-based methods of vessel segmentation | Glaucoma diagnosis | Diabetic retinopathy | To enable multi-disease detection research |

**Abbreviations:** LBC 1936 (Lothian Birth Cohort 1936; hereafter, LBC); ORIGA (Online Retinal Image Database for Glaucoma Analysis and Research); FIVES (Fundus Image VEssel Segmentation); RIM-ONE (Retinal IMage database for Optic Nerve Evaluation); IDRiD (Indian Diabetic Retinopathy Image Dataset); RFMiD (Retinal Fundus Multi-Disease Image Dataset); FOV (field of view); SD (standard deviation). *This number includes images from professional rugby players, a sub-study within Edinburgh, which were used in model development, but not analysed further, partly because OCT was not captured in this group.

*Optic disc annotation*

We annotated the optic disc in 536 images from three datasets (100 ORIGA, 339 PREVENT, 97 LBC). In fundoscopy, it is widely accepted that the disc margin lies at the inner edge of the border tissue, defined as collagenous tissue that arises from the sclera to join Bruch's Membrane, forming a scleral "cuff" or "lip" between the optic nerve head and the choroid, which gives rise to its characteristic appearance as a yellow-white halo/crescent (Figure 3).[19,20] Accordingly, the annotation protocol required dragging the waypoints of a deformable ellipse to the inner edge of the border tissue in full resolution RGB fundus images (Figure 3). Annotation was performed by a single researcher (author SG, a PhD student in retinal image analysis), using custom MATLAB code.

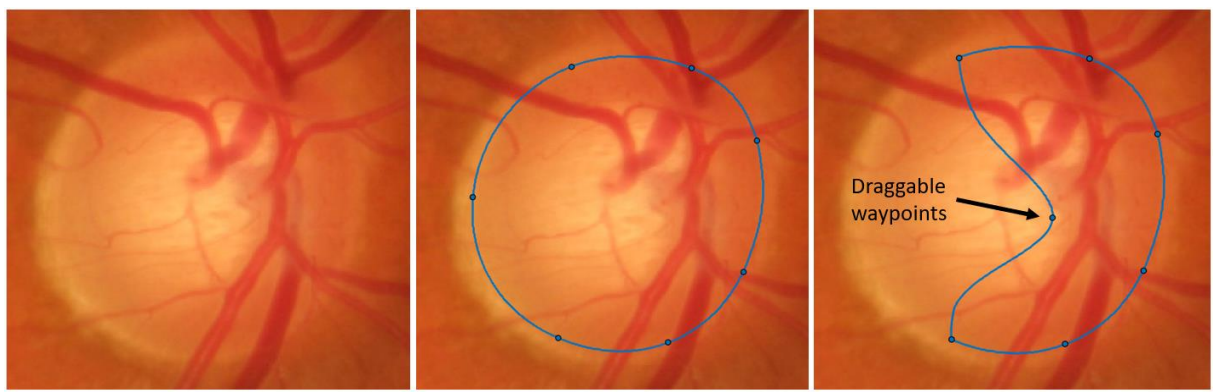

**Figure 3.** Annotation procedure. The annotator loads a full-sized image, and zooms into the optic disc. The user then drags the waypoints of a deformable ellipse to the desired location. Additional waypoints can be added by double-clicking. The entire shape can also be dragged. Performed in MATLAB with custom written code.

To assess inter-annotator agreement, a second researcher (author FY, a PhD student in retinal image analysis, and optometrist) annotated a subset of 100 images (30 ORIGA, 40 PREVENT, 30 LBC) using the same protocol. The agreement metric was mean Intersection over Union (mIoU), calculated as the area of overlap divided by the area of union. mIoU has been shown to be a suitable measure of inter-annotator agreement for medical image segmentation tasks.[21] mIoU between the two annotators was 0.942 (94.2%). Overall, FY annotated a smaller area than SG (mean number of pixels for all images = 75,844 vs. 84,701), however this difference was largely driven by a low agreement on a few images (see histogram in Figure 4, and examples in Figure 5).

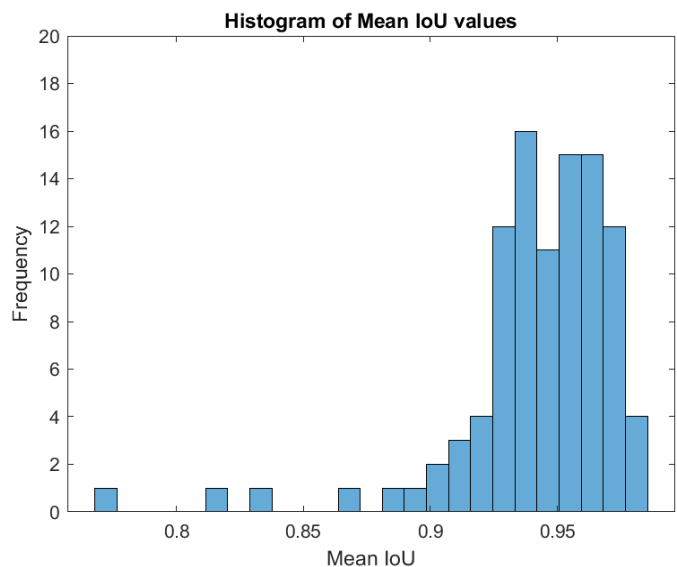

**Figure 4.** Histogram of inter-annotator agreement. **Abbreviations:** IoU (intersection over union).

| Original image | Annotator 1 | Annotator 2 | Overlap |
|---|---|---|---|

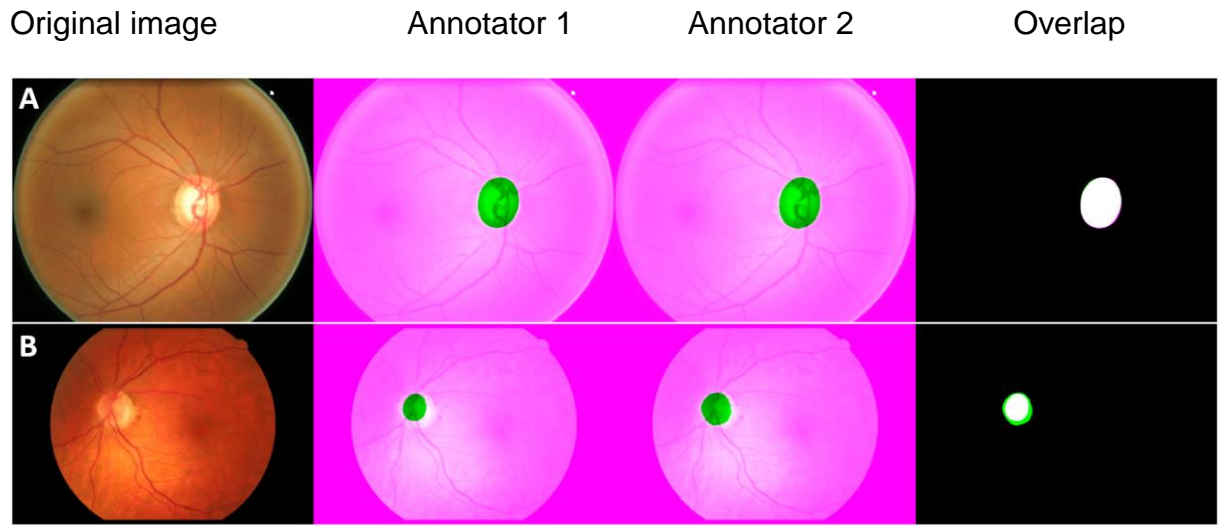

**Figure 5.** Inter-annotator agreement. Row A has a high level of agreement, whereas row B has a low-level agreement. Annotator 1 (SG), Annotator 2 (FY).

*Fovea annotation*

We annotated the fovea in 1,870 images from the LBC and PREVENT datasets (280 PREVENT, 1,590 LBC). The annotation procedure involved dragging a circle with a fixed radius of 150 pixels onto the estimated centre of the fovea to generate a binary fovea map, where pixels inside the circle were labelled as "fovea" and pixels outside labelled as "background". The image presented to the annotator was pre-processed to improve visibility by extracting the green channel of the RGB image and contrast was enhanced by performing contrast limited adaptive histogram equalization.[22] When the fovea was not visible (i.e., due to poor illumination), the annotator estimated its position relative to the vessel arc of the central arcades, and disc. If neither the vessel arc nor the disc was visible, the image was rejected. According to this protocol, 33 of 1,623 images were rejected from the LBC, and none were rejected from PREVENT. Annotation was performed by a single researcher (author SG) using custom MATLAB code. No second annotator was used. The annotation and procedure are visualized in Figure 6.

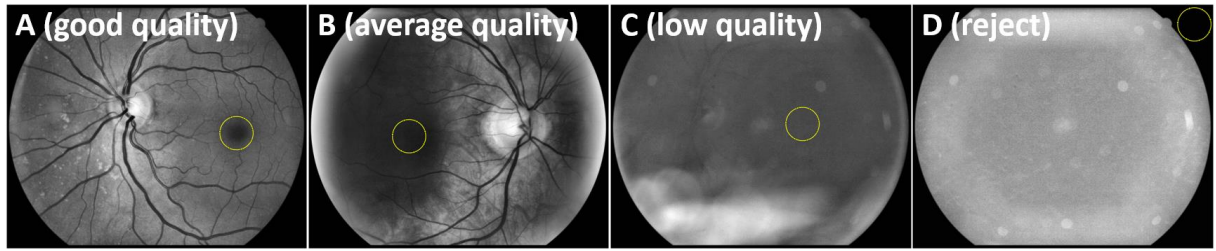

**Figure 6.** Fovea annotation procedure. Panel A shows a good quality image, with the fovea clearly visible. Panel B has an area of low illumination over the macula, but the fovea can still be estimated. Panel C has very low illumination and blur across the image, however the optic disc and vessel arc are still visible, allowing the fovea to be estimated. In panel D, neither the vessel arc nor the optic disc is visible – fovea estimation not possible.

*CNN architecture and computing platform*

Based on a 2022 survey of deep learning-based image segmentation,[23] we selected Google's DeepLabv3+ architecture,[24] which was the best performing network for image segmentation among the networks reviewed. DeepLabv3+ incorporates an encoding and decoding phase. The encoder/decoder model has been described elsewhere.[23,24] Briefly, in the encoding phase, information from the input image is extracted and compressed into a feature representation using a 'backbone' convolutional neural network. The decoder then takes this as input to reconstruct the initial representation. The goal of such encoder-decoder architecture is for the model to learn a useful representation of the image. The result is accurate segmentation along object boundaries.[24] DeepLabv3+ can take one of several backbone architectures, including MobileNetv2[25] and Xception.[26] During experimentation, we found that MobileNetv2 produced the best results. Accordingly, we used DeepLabv3+ with a MobileNetv2 backbone pretrained on ImageNet[27] for all models except vessel segmentation, for which we selected Xception. The values of relevant parameters varied with tasks and are given in the next section.

All models were trained in MATLAB (*version R2022b*, Natick, Massachusetts: The MathWorks Inc) using the Deep Learning Toolbox, on a Dell 7820 machine, fitted with an Intel Xeon Silver CPU and a NVIDIA Quadro RTX 5000 GPU, running Windows 10.

*Optic disc localization*

The optic disc localisation network was trained on 536 images from three datasets (100 ORIGA, 339 PREVENT, 97 LBC). In pre-processing, we resized all images and their corresponding labels to 650 × 650 pixels, allowing the whole image to be processed by the network. We then split images into training, validation, and test sets, with ratio 80/10/10, yielding 429 images for training, 54 for validation, and 53 for testing. We used the Adam optimizer, the learning rate was constant at 0.0001, and the batch size was 4. During training, validation was carried out after every 100 iterations (approximately every epoch). We applied the following data augmentations to the training images to enhance generalisation to unseen data: addition of random colour jitter (brightness = 0.3, contrast = 0.3, saturation = 0.3), scaling (between a factor of 0.8 and 1.3) and rotation (between -30 and 30˚). To prevent overfitting, we finalized training if the validation loss stopped decreasing or was equal to the previous 10 losses ("validation patience").

The aim of the task was to locate the approximate centre of the disc. There were far fewer pixels labelled as "disc" than pixels labelled as "background" (ratio = 1/65), leading to an imbalanced training set. With common metrics, a model would perform quite well if all pixels were simply labelled as "background". To account for imbalance, we replaced the classification layer with a class weights layer, based on the class distribution of the full image set.

The convergence criterion was met after 3,700 iterations (during the 28th epoch). The final validation accuracy was 99.16%. After post-processing (removing the smallest object where multiple objects were detected), we computed the mean Euclidean distance in pixels (mED) between the ground truth and prediction based on the central points (Figure 7). The mED in the test set was 2.06 pixels (SD = 1.21), which expressed as a percentage of disc size (major axis length) was 2.02% (SD = 1.2%).

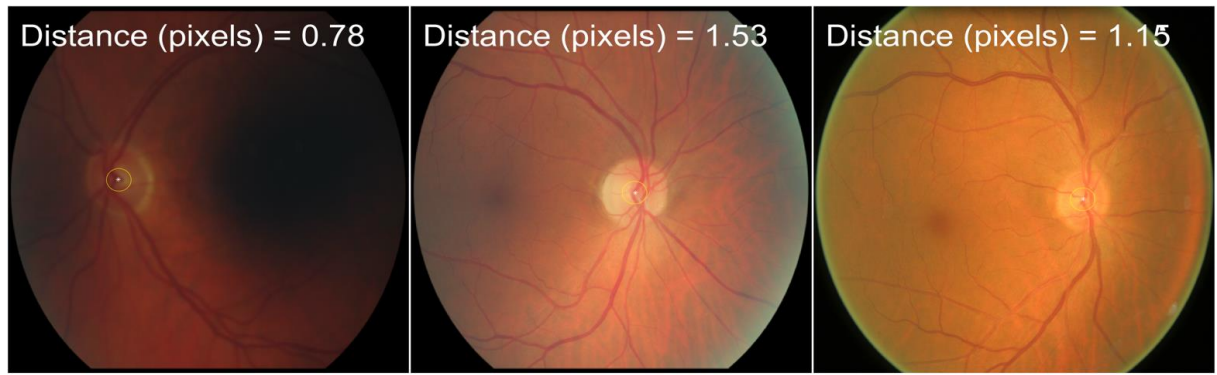

**Figure 7.** Disc localization results on the test set. The ground truth is represented by a circle, and the prediction by an asterisk.

*Optic disc segmentation*

The optic disc segmentation network was trained on 536 images from three datasets (100 ORIGA, 339 PREVENT, 97 LBC). Input to the network was a 650 × 650 × 3 RGB image, cropped around the disc, plus its corresponding ground truth segmentation. We split images into training, validation, and test sets, with ratio 80/10/10 respectively. We used the Adam optimizer, the learning rate was constant at 0.0001, and the batch size was 4. During training, validation was carried out after every 100 iterations (approximately every epoch). Augmentations were applied as before, but with additional levels of random colour jitter (brightness = 0.6, contrast = 0.6, saturation = 0.6) and scaling (between a factor of 0.4 and 1.6). As before, class imbalance was high, albeit less than before (ratio ≈ 1/4.17), which we accounted for in the same way. The validation patience was 20. The convergence criterion was met after 16,400 iterations (during the 123$^{rd}$ epoch). The final validation accuracy was 98.79%. The mIoU for the test set (53 images) was 0.959 (95.9%). Two examples from the test set are presented in Figure 8.

| Original | Ground Truth | Prediction | Overlay |
|----------|--------------|------------|---------|

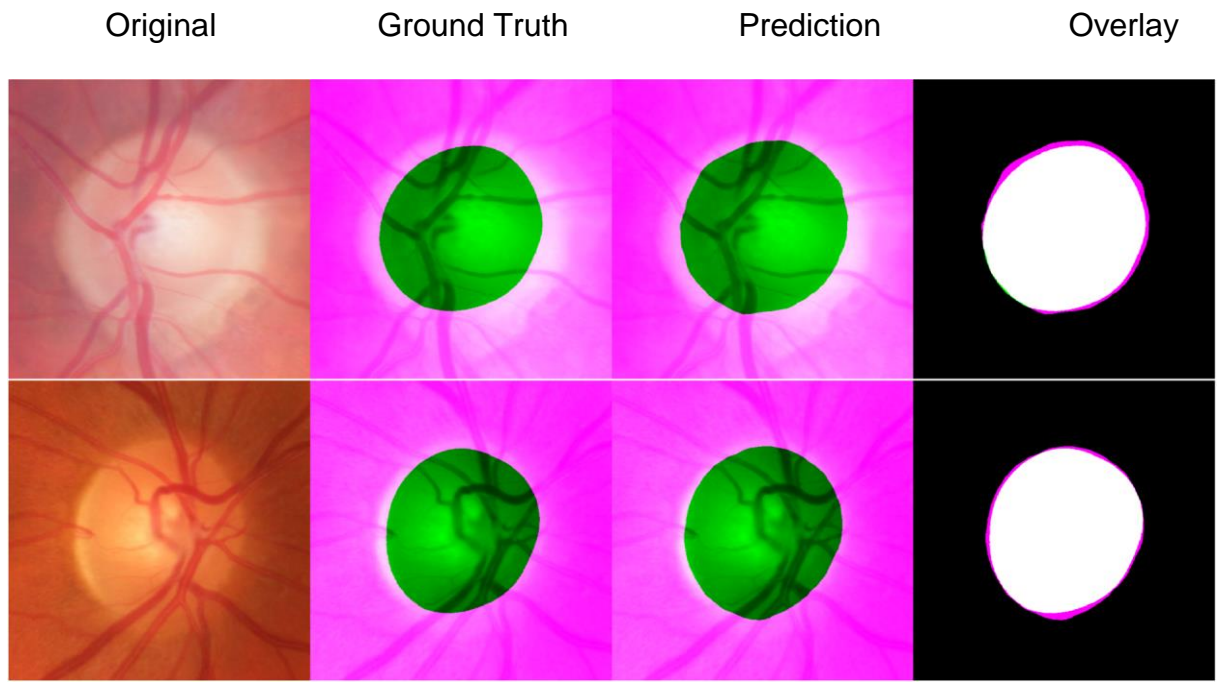

**Figure 8.** Optic disc boundary predictions from the test set.

We carried out external testing on IDRiD and RIM-ONE. Prior to generating predictions for IDRiD, each input image was resized, while maintaining its aspect ratio, to the median height of the training images. This ensured that when the image was cropped around the disc, the ratio of disc to background was approximately similar to that of a typical image from the training set. We carried out this step to improve generalizability.

The RIM-ONE dataset contains images that are already cropped around the disc; therefore, we skipped the disc localization stage. Our main performance metric was mIoU, however, to enable better comparison with other work, we also provide mean accuracy (mAcc), defined as TP / (TP + FN), where TP is a true positive and FN is a false negative. We provide a direct comparison with the state-of-the-art for IDRiD, however for RIM-ONE we applied our network to their most recent data release (release 3). Other networks cited here have only been applied to release 1, therefore comparative results are indicative only. On IDRiD, our model achieved a mIoU of 0.891, which beat the current state-of-the-art (0.845). On RIM-ONE, we achieved a mIoU of 0.926 for non-glaucomatous eyes, and 0.907 for glaucomatous eyes. A comparative summary is shown in Table 2.

**Table 2.** Comparison of our OD segmentation model to state-of-the-art.

| Method | Year | IDRiD (81 images) | | RIM-ONE V3 (non-glaucoma; 313 images) | | RIM-ONE V3 (glaucoma; 172 images) | | RIM-ONE V1 (169 images) | |
|---|---|---|---|---|---|---|---|---|---|
| | | mAcc | mIoU | mAcc | mIoU | mAcc | mIoU | mAcc | mIoU |
| ResNet + Unet | 2020 | - | - | - | - | - | - | - | 0.925 |
| DRNet | 2021 | 0.997 | 0.845 | - | - | - | - | 0.962 | 0.901 |
| Ours | 2022 | 0.899 | **0.891** | 0.958 | **0.926** | 0.947 | 0.907 | - | - |

**Abbreviations:** mAcc (mean accuracy), mIoU (mean intersection over union). ResNet + Unet,[28] DRNet.[29] Notes: Best score is in bold, second best is underlined.

*Fovea localization*

The fovea localisation network was trained on 1,870 images from the LBC and PREVENT datasets (280 PREVENT, 1,590 LBC). In pre-processing, we resized each image to 224 × 224 pixels. Unlike other features, the fovea is often not visible, however its location can be inferred from its position relative to other salient features, including the vessel arc of the central arcades, and the large dark patch covering the macula, which is common in poorly illuminated images. We hypothesized that substantially reducing image size would force the network to focus on these features. Therefore, input to the network was a 224 × 224 × 3 RGB image and its corresponding label. Images were split into training, validation, and test sets, with ratio 70/10/20 respectively, yielding 1,309 images for training, 187 for validation, and 374 for testing. We chose a relatively large (20%) test set for this task as the focus was on generalisation. We used the Adam optimizer, the learning rate was constant at 0.0001, and the batch size was 64. During training, validation was carried out after every 20 iterations.

We manually stopped training after 1,167 iterations (during epoch 41), when we observed that the model was no longer improving. In postprocessing, we removed the smallest object, where multiple objects were detected. As with the optic disc localisation network, we measured performance in the test set by calculating the Euclidian distance between the ground truth and prediction based on their central points. mED was 3.77 (SD = 2.71). Expressed as a percentage of image height (224 pixels; disc size was not consistently available due to image quality), mED was 1.68% (SD = 1.11%). Several results from challenging images in the test set are presented in Figure 9.

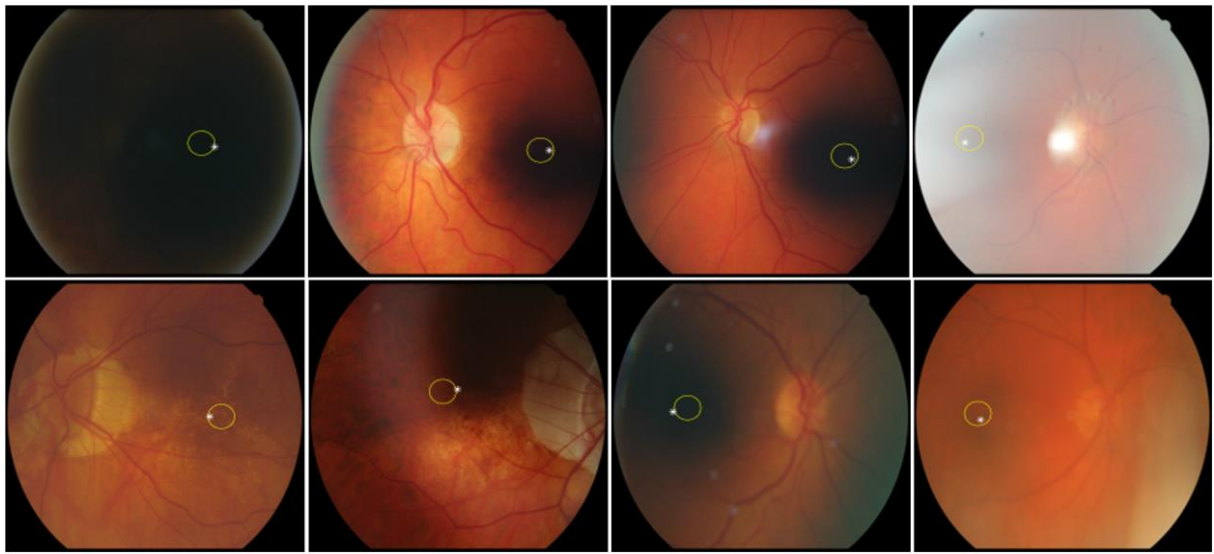

**Figure 9.** Fovea estimations in challenging images from the test set. The yellow circle represents the ground truth, and the asterisk is the prediction.

We carried out external testing on IDRiD (103 images). In addition to providing mED, we evaluate performance with the 1R criterion,[30] which refers to the radius of the optic disc. The 1R grid is centered on the fovea, and a score of 1 is given if the predicted coordinates lie within a given region (Figure 10). mED was 64.38 (SD = 76.43), and medianED was 38.71. 95.15% of predictions fell within 1R, 86.41% within 0.5 R, 71.84% within 0.25 R, and 4.85% fell outside 1R (failed). The current state-of-the-art for IDRiD is mED 41.87.[29]

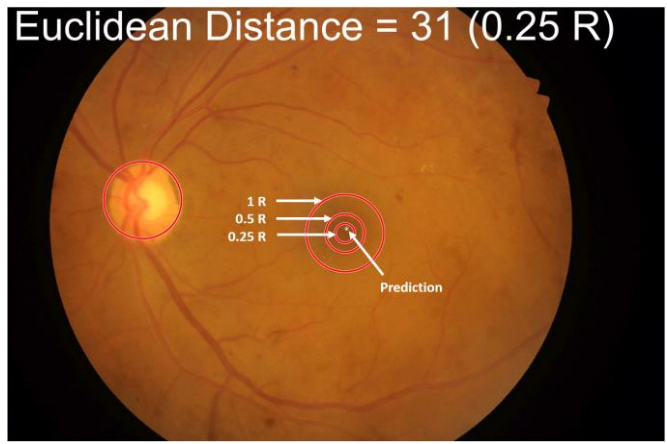

**Figure 10.** Both Euclidean distance and the 1R criterion were used to evaluate performance in the fovea detection network. The circular 1R grid is centered on the fovea. Image shown is from IDRiD.

*Vessel segmentation*

We trained the vessel segmentation network on 800 images from the FIVES dataset. In pre-processing, we used the optic disc localisation network, described earlier, to crop each image and its counterpart vessel mask to 650 × 650 pixels centred on the disc. Therefore, input to the network was a 650 × 650 × 3 RGB image and its corresponding label. We split images into training, validation, and test sets, with ratio 70/15/15, resulting in 560 images for training, 120 for validation, and 120 for testing. Unlike our previous networks, we used Xception[26] as the backbone, as it generated more accurate segmentations during experimentation. We used the Adam optimizer, the learning rate was 0.0001, which we set to decrease by a factor of 0.1 in a piecewise manner every 5 epochs. The batch size was 4, and validation was carried out every 100 iterations. Augmentations applied were identical to those used for disc localization.

We manually stopped training after 3,561 iterations (during epoch 6), as the model had converged. The final validation accuracy was 97.2%. Mean accuracy on the test set was 95.43%, mIoU was 0.88. An example of the ground truth and automatic result is presented in Figure 11.

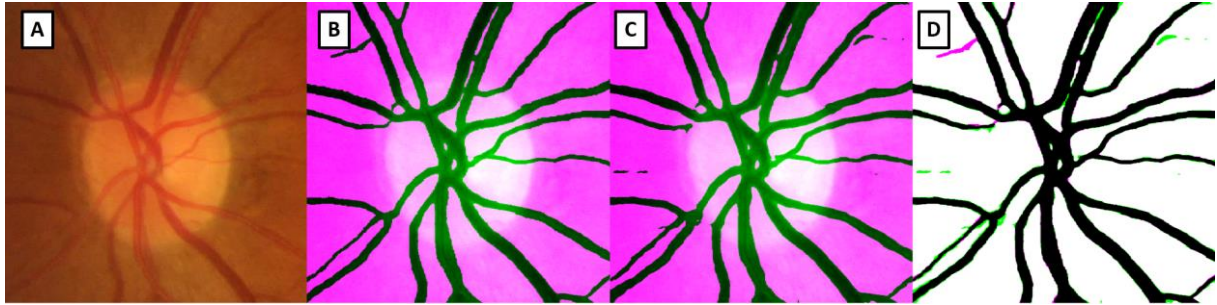

**Figure 11.** Vessel segmentation. A) Input to the network, B) ground truth, C) automatic result, and D) superposition of false-colour image of ground truth and automatic result (false negative = green, false positive = magenta).

*Generating pallor measures*

To calculate pallor, the software took as input full-size colour fundus photographs. In pre-processing, a 300-pixel border of zeros was added to the left and right sides of the image, and the whole image was resized to the median height of the training images (2,166 pixels), while maintaining its aspect ratio. Adding the border prevented a cropping failure when the disc was close to, or on the border, and resizing helped the model generalize by ensuring that the input image was approximately equivalent in size to the images the network was trained on.

The disc localization network was used to locate the disc centre. Then, the image was cropped to a size of 650 × 650 pixels, with the disc at the centre. Next, disc segmentation was performed on the cropped image. Post-processing of the predicted disc boundary involved keeping the largest object (where multiple objects were detected), filling holes (where holes were detected), and smoothing edges.

Edge smoothing consisted of three stages, 1) morphological opening with a disc-shaped structuring element of radius 75, 2) blur with a 2D convolution, 3) re-threshold to a value of 0.5.[*]

We defined the measurement region as starting at the inner edge of the border tissue and extending a fixed distance of 30 pixels inwards. We chose this distance through direct observation as a balance between capturing as much of the neuro-retinal rim (NRR) as possible while avoiding the cup. We defined the control region as starting at the outer border of the cropped image and extending a fixed distance of 50 pixels inwards. We chose this distance through direct observation as a compromise between capturing as much of the retina as possible while avoiding the disc and any atrophy. Vessels were detected in the cropped image and excluded (vessel pixels replaced with zero) from both the measurement region and the control region.

We then divided the measurement region into zones in accordance with the Heidelberg system for assessing pRNFL thickness. Specifically, the intersection of the optic disc-fovea axis and the measurement region took a value of zero degrees. The temporal zone then extended from 45° to -45°, the temporal inferior from 45° to 90° and so on. The papillomacular bundle is a special case of the temporal zone, extending from 15° to -15°.

Finally, we calculated pallor based on the ratios of red and green pixel intensities[6,7,10,31,32] between the measurement and control region. Specifically, we divided the mean of the green channel in the measurement zone by the mean of the red channel in the same zone. The result was then divided by the same measurement in the control region, except using the medians instead of the means. The result was a measure of pallor within each eye, for each zone.

---

[*] Edge smoothing algorithm taken directly from MATLAB user "Image Analyst", available at
https://uk.mathworks.com/matlabcentral/answers/380687-how-to-smooth-rough-edges-along-a-binary-image

*Statistical analysis*

Data from one eye is correlated with data from the fellow eye.[33] Ying *et al.*[34] showed that ignoring this inter-eye correlation in standard regression models can lead to spurious conclusions. The authors suggest that linear mixed effects modelling with the eye as the unit of analysis should be used. Accordingly, we modelled a random intercept for each person and eye. We adjusted p-values for multiple comparisons with the False Discovery Rate (FDR) procedure, which accounts for correlation between measurements. Statistical analysis was performed in R (version 4.2.1; www. R-project.org) using the lme4 and LmerTest packages.

To select covariates, we follow the "dijunctive cause criterion",[35] which states that covariates should be added if they are causes of the exposure *or* outcome, or causes of both. Accordingly, the two retinal covariates are disc size and image brightness.

Disc size: The NRR must accommodate between 0.9 and 1.5 million RGC axons. In a large disc, the axons can spread out, whereas in a small disc they are more compact. This means that, theoretically, the larger the disc, the paler it will appear, and vice versa. Indeed, we found a correlation between disc size and global pallor ($R^2 = 0.11$).

Brightness: We defined brightness as the median of all pixels inside the control region, after converting to greyscale, and removing vessels. While the software controls for image brightness within each eye, light reflectance in the fundus is known to vary depending on the which part of the retina it strikes.[36]

Specifically, the proportion of light reflected from the NRR and the background fundus may not remain constant with the level of light entering the pupil though the camera flash. Indeed, we found a correlation between brightness in the control region and global pallor ($R^2 = 0.12$).

*Interocular Variability*

We assessed interocular differences in pallor for each zone and propose a new measure "Inter-ocular Pallor Variability" (IoPV), defined as the sum of absolute differences from all six zones.

*Detecting pallor in the RFMiD dataset*

We tested the results of our software on 92 images from the RFMiD dataset (images taken from the training set), half of which had been labelled as "optic disc pallor", and half as "disease risk = 0" by ophthalmologists from the RFMiD group. To assess group differences, we performed unpaired two-sampled Wilcoxon tests.

*Testing the software for robustness to camera system, format, and resolution*

Among different countries, clinics, and research institutions, there is considerable heterogeneity in the technical aspect of retinal fundus images, including i) the camera (e.g., Topcon, Canon), ii) image resolution and image size, and iii) file format (e.g., JPG, PNG, TIFF). To test the resilience of our system to these factors, a dataset was constructed containing images captured by various imaging systems, with different resolutions and formats. In addition, details pertaining to the field of view (FOV), centering protocol, and dilation were also noted. We chose 5 sets of 10 images from a total of 4 datasets (G1020[37], MESSIDOR[38] PREVENT, and REFUGE[39]). All these datasets, except for PREVENT, are publicly accessible. The task focus was on whether images could be successfully processed, not on how the software copes with images of varying levels of quality. Accordingly, we selected images with sufficient quality (broadly even illumination, free of major pathology). We judged the results by visual inspection, according to whether the software correctly a) located the fovea, b) located and segmented the disc, c) rotated the image along the optic disc-fovea axis, and d) segmented the vessels. We also recorded computation time for each batch to assess whether processing time differed by dataset.

*Developing a set of automatic rejection criterion*

If there was insufficient information in an image to localise the disc or fovea, the image failed at the stage of processing. These images were usually very over- or under-exposed (i.e., near totally white or black), or contained excessive blur. However, in most cases, the software processed the image, even when quality was very low. To enable processing on large datasets, we aimed to develop a set of criteria through which such images could be automatically rejected. That is, although the images have been successfully processed, they are clearly not suitable for further analysis. For this task we used the LBC dataset, which contains images of varying levels of quality. We propose two automatic rejection criteria: disc eccentricity and control region brightness.

*1) Eccentricity* is the ratio of the distance between the centre of an ellipse fitted onto the disc, and the major axis length, where 0 is a circle, and 1 is a line.

*2) Control region brightness* is the median of all pixels inside the control region, after converting to greyscale, and removing vessels.

By visual inspection, we aimed to develop conservative thresholds that would reject only the poorest quality images, or cases in which the software clearly failed for another reason (e.g., severe pathology).

**Results**

The software takes less than 3 seconds to process a single image and outputs several key visualizations (Figure 12) alongside tabular data.

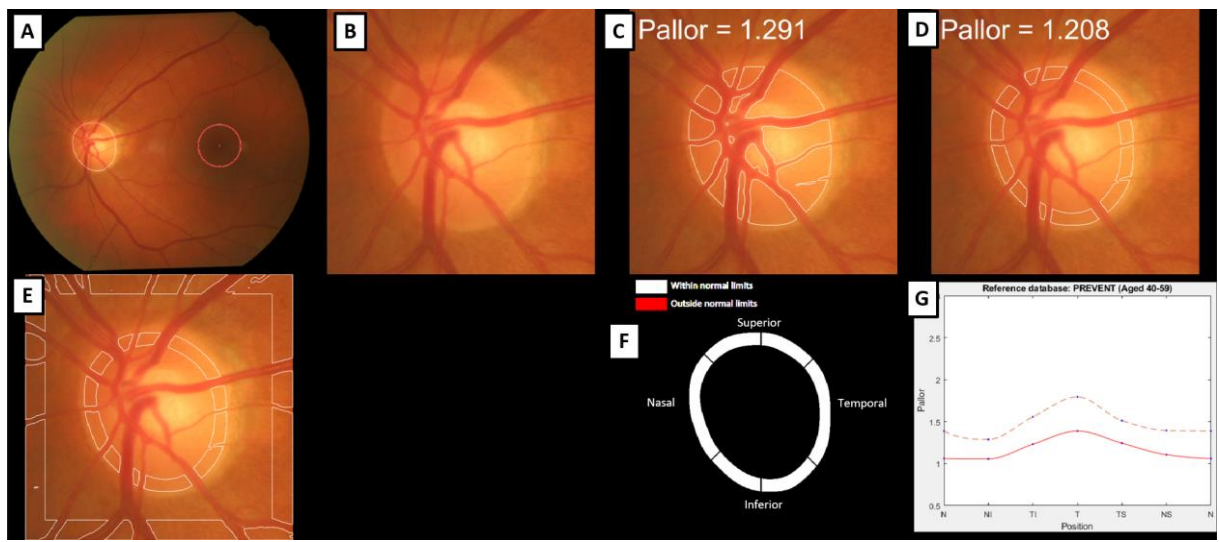

**Figure 12.** Core visualizations of the software. A) Disc and fovea localization are used to rotate the image along the optic disc-fovea line; B) Cropped optic disc; C) Segmented disc excluding vessels; D) Measurement region excluding vessels; E) measurement and control region (outer square) excluding vessels;  F) Alert system (region lights up red if a limit is exceeded); G) Dashed line represents one standard deviation above the mean of all participants in PREVENT, red line is the current participant.

*Quality control and sample derivation*

The sample derivation is illustrated in Figure 13. After quality control, concurrent fundus images and OCT scans were available for 118 participants (226 eyes). Three fundus images were rejected due to segmentation error and low illumination (author SG; visual inspection; Supplementary Figure 1), and 13 OCT scans were excluded for reasons including clipping (4 images), improper centering (4 images), high myopia (≤ −5 diopters; 2 images), poor segmentation (3 images), poor

illumination (1 image), and signs of pathology (5 images). Pathologies in OCT

included epiretinal membrane, excessive peripapillary atrophy, and tilted discs. OCT

quality control was carried out by CH (author), an ophthalmic imager and analyst, via manual inspection of the images through the Heidelberg platform.

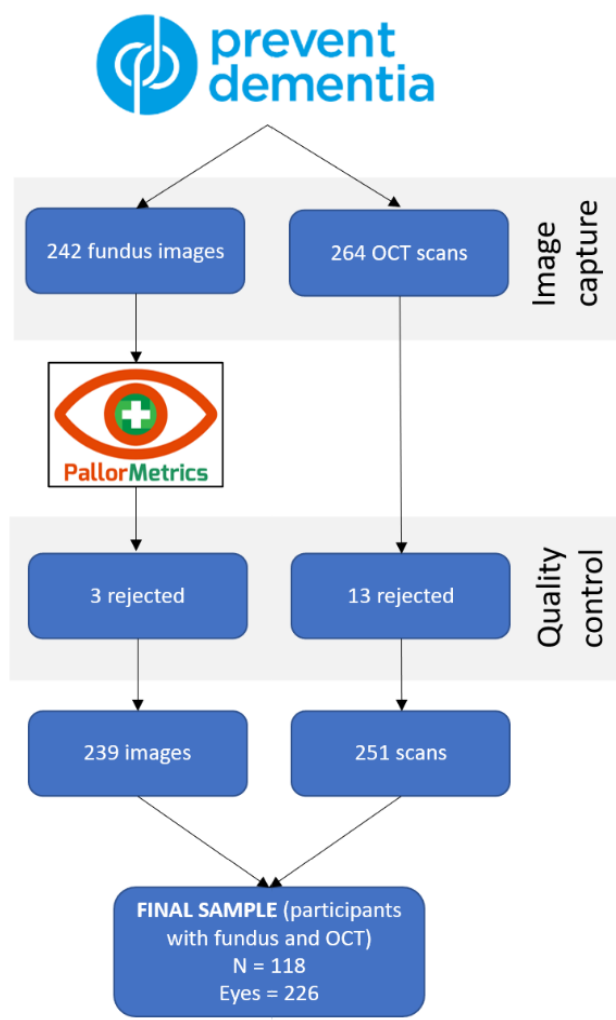

**Figure 13.** Sample derivation flowchart.

*Summary statistics*

Image characteristics are summarized in Table 3, alongside basic

demographics. Histograms showed that pallor was normally distributed

(Supplementary Figure 2).

**Table 3.** Demographics, covariates, pallor, and pRNFL thickness in microns by zone and eye.

| | Eye | | |
|---|---|---|---|
| | **Left** | **Right** | **Overall** |
| N (participants) | | | 118 |
| N (eyes) | (N=112) | (N=114) | (N=226) |
| Age (years) | 51.5 (± 5.60) | 51.4 (± 5.67) | 51.4 (± 5.63) |
| Sex (female) | 68 (60.7%) | 68 (59.6%) | 136 (60.2%) |
| Disc area | 75,300 (± 16,400) | 72,400 (± 14,500) | 73,900 (± 15,500) |
| Luminance | 93.1 (± 19.5) | 90.1 (± 18.8) | 91.6 (± 19.1) |
| **Temporal** | | | |
| Pallor | 1.62 (± 0.24) | 1.54 (± 0.22) | 1.58 (± 0.23) |
| RNFL | 69.3 (± 13.8) | 73.8 (± 13.1) | 71.6 (± 13.6) |
| RNFL missing | 5 (4.5%) | 6 (5.3%) | 11 (4.9%) |
| **Temporal Inferior** | | | |
| Pallor | 1.42 (± 0.19) | 1.34 (± 0.19) | 1.38 (± 0.19) |
| RNFL | 139 (± 19.7) | 141 (± 22.7) | 140 (± 21.2) |
| RNFL missing | 5 (4.5%) | 6 (5.3%) | 11 (4.9%) |
| **Nasal Inferior** | | | |
| Pallor | 1.16 (± 0.14) | 1.12 (± 0.14) | 1.14 (± 0.14) |
| RNFL | 113 (± 21.9) | 112 (± 22.1) | 113 (± 22.0) |
| RNFL missing | 5 (4.5%) | 6 (5.3%) | 11 (4.9%) |
| **Nasal** | | | |
| Pallor | 1.25 (± 0.16) | 1.19 (± 0.15) | 1.22 (± 0.18) |
| RNFL | 74.7 (± 16.2) | 76.9 (± 18.9) | 75.8 (± 17.6) |
| RNFL missing | 6 (5.4%) | 7 (6.1%) | 13 (5.8%) |
| **Nasal Superior** | | | |
| Pallor | 1.27 (± 0.16) | 1.20 (± 0.15) | 1.23 (± 0.16) |
| RNFL | 111 (± 18.9) | 98.6 (± 18.8) | 105 (± 19.8) |
| RNFL missing | 6 (5.4%) | 7 (6.1%) | 13 (5.8%) |
| **Temporal Superior** | | | |
| Pallor | 1.36 (± 0.19) | 1.30 (± 0.18) | 1.33 (± 0.19) |
| RNFL | 135 (± 17.3) | 135 (± 16.9) | 135 (± 17.0) |
| RNFL missing | 5 (4.5%) | 6 (5.3%) | 11 (4.9%) |
| **PMB** | | | |
| Pallor | 1.68 (± 0.25) | 1.59 (± 0.23) | 1.63 (± 0.25) |
| RNFL | 53.9 (± 13.1) | 55.6 (± 9.04) | 54.7 (± 11.2) |
| RNFL missing | 6 (5.4%) | 9 (7.9%) | 15 (6.6%) |
| **Global** | | | |
| Pallor | 1.41 (± 0.18) | 1.34 (± 0.17) | 1.37 (± 0.18) |
| RNFL | 98.1 (± 8.03) | 98.2 (± 8.12) | 98.2 (± 8.06) |
| RNFL missing | 6 (5.4%) | 7 (6.1%) | 13 (5.8%) |
| **Nasal/Temporal Ratio** | | | |
| Pallor | 0.77 (± 0.07) | 0.78 (± 0.08) | 0.78 (± 0.07) |
| RNFL | 1.13 (± 0.34) | 1.08 (± 0.36) | 1.10 (± 0.35) |
| RNFL missing | 7 (6.3%) | 10 (8.8%) | 17 (7.5%) |

| Whole Disc | | | |
|---|---|---|---|
| Pallor | 1.46 (± 0.201) | 1.39 (± 0.185) | 1.43 (± 0.196) |

**Abbreviations:** RNFL (retinal nerve fibre layer); PMB (papillomacular bundle). All values are Mean (SD; standard deviation) or N (%).

Pallor was highest temporally, and lowest nasally. Mean pallor was lower when considering all zones in the measurement region (global pallor; 1.37, SD = 0.18) than when considering the entire disc (mean = 1.43, SD = 0.2). In all measurement zones, pallor was numerically higher in the left eye compared with the right eye. pRNFL was thickest in both polar zones (superiorly, and inferiorly), which was in accordance with typical findings.[40] Unlike pallor, pRNFL was not systematically different between the eyes. Boxplots of pallor and pRNFL by zone and eye are presented in Figure 14.

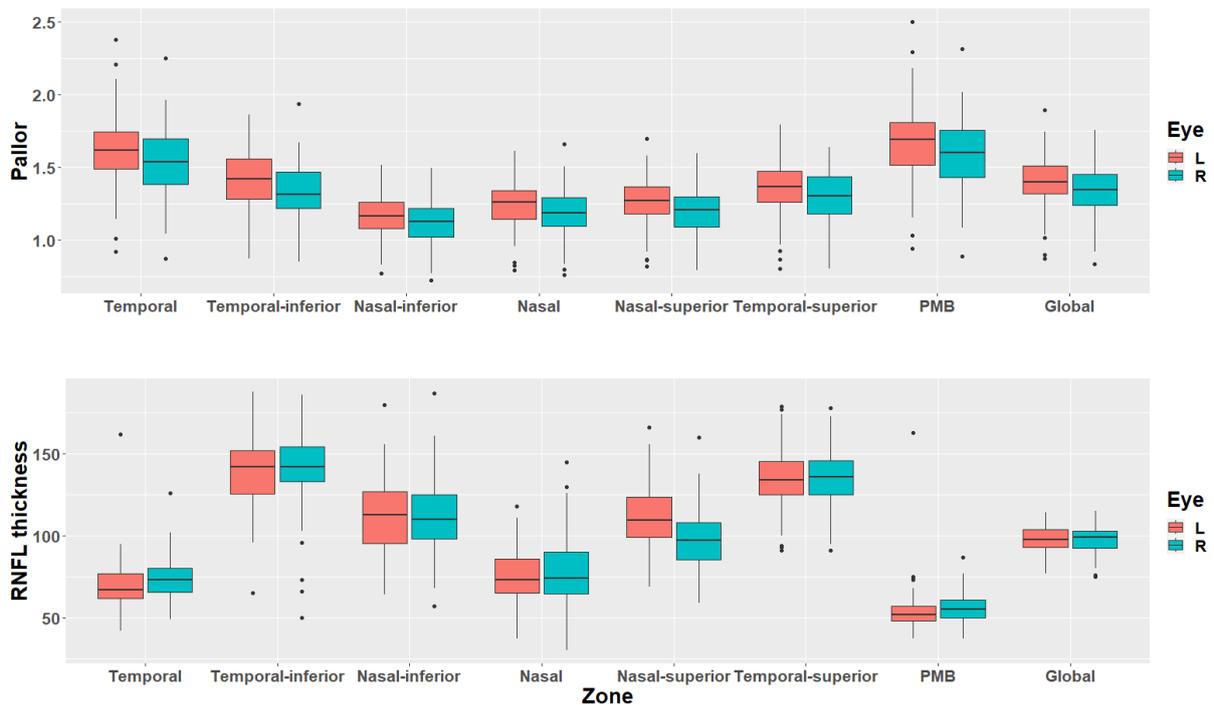

**Figure 14.** Boxplots representing pallor and pRNFL thickness values by zone and eye. N = 118 (114 right eye, 112 left eye).

*Associations between pRNFL thickness and pallor*

After adjusting for age, sex, disc area, control region brightness, and multiple comparisons, we observed statistically significant associations between pRNFL thickness and pallor globally ($\beta$ = -9.81 (SE = 3.16), p < 0.05), in the temporal inferior zone ($\beta$ = -29.78 (SE = 3.32), p < 0.01), and with the nasal/temporal ratio ($\beta$ = 0.88 (SE = 0.34), p < 0.05). Pallor in the measurement region was more discriminative than pallor measured in the whole disc ($\beta$ = -8.22 (SE = 2.92), p < 0.05). We also found an association between pRNFL thickness and pallor in the temporal-superior zone ($\beta$ = -17.29 (SE = 7.83), p < 0.05), however this did not survive correction for multiple comparisons. Results are summarised in Table 4.

**Table 4**. Linear mixed effects regression models of pRNFL thickness predicted by pallor in equivalent zones. Random intercepts are modelled for each subject and eye. Models are adjusted for age, sex, disc area, and image brightness.

| Zone | Coefficients | | |
|---|---|---|---|
| | *β (SE)* | *P* | *P (adjusted for FDR)* |
| Global | **-9.81 (3.16)** | **0.002**\*\* | **0.011**\* |
| Temporal | -2.89 (4.68) | 0.538 | 0.724 |
| Temporal-inferior | **-29.78 (8.32)** | **0.000**\*\*\* | **0.004**\*\* |
| Nasal-inferior | 5.92 (13.09) | 0.652 | 0.724 |
| Nasal | -4.7 (9.2) | 0.610 | 0.724 |
| Nasal-superior | -7.92 (9.85) | 0.422 | 0.704 |
| Temporal-superior | **-17.29 (7.83)** | **0.028**\* | 0.057 |
| PMB | -0.03 (3.72) | 0.994 | 0.994 |
| Global pRNFL ~ pallor in whole disc | **-8.22 (2.92)** | **0.005**\*\* | **0.018**\* |
| Nasal-temporal ratio | **0.88 (0.34)** | **0.011**\* | **0.028**\* |

**Abbreviations:** PMB (Papillomacular Bundle); pRNFL (peripapillary retinal nerve fibre layer); SE (standard error); FDR (false discovery rate). \*P<0.05, \*\*p<0.01, \*\*\*p<0.001.

*Interocular Pallor Variability (IoPV)*

In the PREVENT dataset, data from both eyes was available for 108 participants. For global pallor, we measured a mean unit difference of 0.1 (SD = 0.07) between the eyes. To put this into context, global pallor ranges from 0.87 to 1.9. This means that although interocular difference is evident, measurements from one eye in a person are broadly similar to measurements from the fellow eye. We also observed differences between zones, for example, the biggest difference between left and right eyes was observed in the temporal region (mean = 0.13, SD = 0.1). The general pattern of pallor being high temporally and low nasally was preserved between the eyes (Figure 15). In Figure 16 (right), pallor is higher in the right eye than the fellow eye in the nasal zone, but higher in the left eye than the fellow eye in the temporal zone. To capture this zone-to-zone variability, we take the sum of absolute differences from all six zones. By contrast, in Figure 16 (left), pallor is higher in all zones in the left eye.

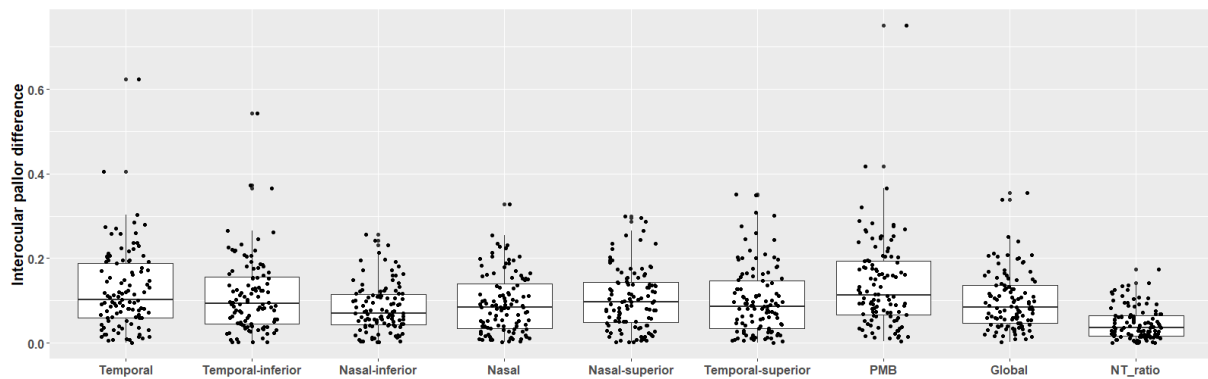

**Figure 15.** Boxplots representing the difference in pallor between the eyes of a participant (N=108).

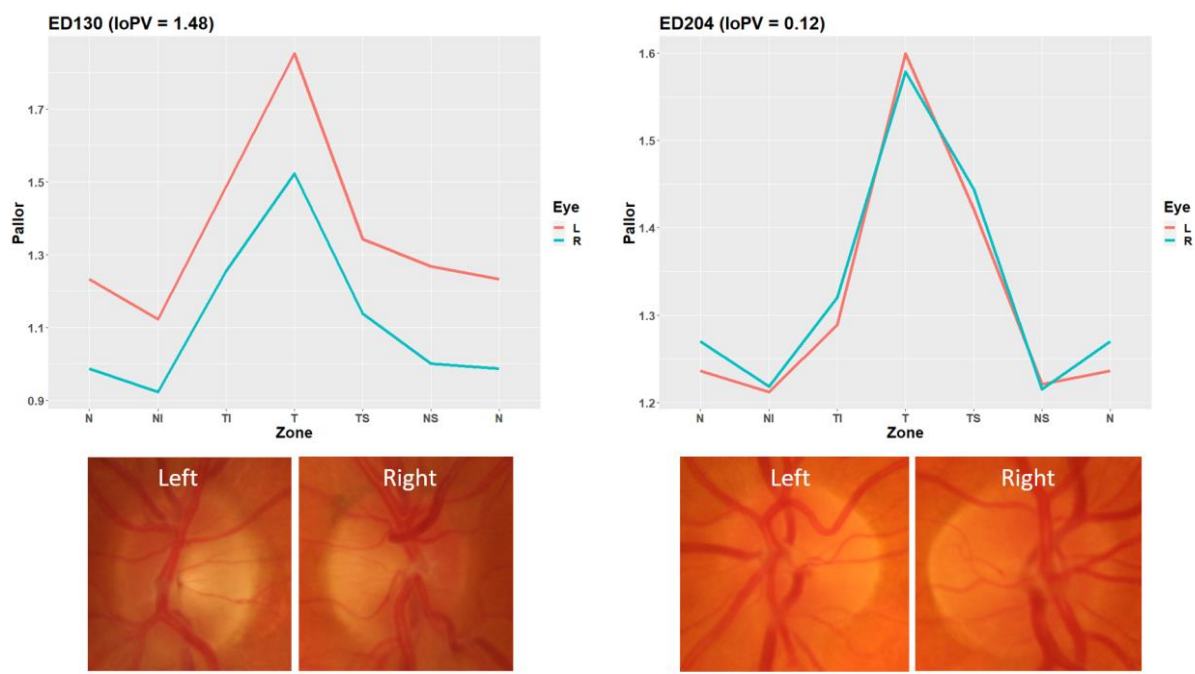

**Figure 16.** Parallel plots from two participants showing interocular differences in pallor by zone (nasal pallor is repeated on either side of each plot for asthetics). **Abbreviations**: IoPV (interocular pallor variability); N (nasal); T (temporal); I (inferior); S (superior).

*Assessing pallor in the RFMiD dataset*

Out of 92 images (46 pallor, 46 controls) in the RFMiD dataset (training set), which contained a patient group (diagnosed optic disc pallor), and healthy controls, the fovea localisation module failed in one patient image, and was subsequently rejected, despite accurate disc segmentation. Accordingly, analysis was carried out on 45 images labelled as pallor, and 46 healthy controls.

Predicted pallor was substantially higher in the patient group compared to the control group for all zones. For example, mean global pallor in the control group was 0.98 (SD = 0.09) compared with 1.23 (SD = 0.14) in the patient group, and this difference was statistically significant (Wilcoxon unpaired signed rank test: W = 208, $p < 10^{-11}$, R = 0.73; Table 5). There was no evidence of significant difference in the nasal/temporal ratio between the groups; control group mean 0.9 (SD = 0.07), patient group mean 0.9 (SD = 0.05), reflecting the diffuse nature of pallor identified in the images. One example from each group is presented visually in Figure 17.

**Table 5.** Unpaired Wilcoxon signed rank test results comparing eyes labelled as having pallor vs controls in the RMFiD dataset. Note: We chose not to correct for multiple comparisons here due to the very low p-values.

| Zone | Group (Mean, SD) | | Wilcoxon test | | |
|---|---|---|---|---|---|
| | Control | Pallor | W | p-value | R (effect size) |
| Temporal | 1.06 (0.12) | 1.32 (0.15) | 195 | $< 10^{-10}$ | 0.7 |
| Temporal-inferior | 0.95 (0.09) | 1.2 (0.13) | 118 | $< 10^{-12}$ | 0.76 |
| Nasal-inferior | 0.91 (0.09) | 1.15 (0.13) | 143 | $< 10^{-11}$ | 0.74 |
| Nasal | 0.95 (0.1) | 1.18 (0.14) | 161 | $< 10^{-11}$ | 0.73 |
| Nasal-superior | 0.93 (0.09) | 1.14 (0.13) | 176 | $< 10^{-11}$ | 0.72 |
| Temporal-superior | 0.97 (0.1) | 1.24 (0.15) | 151 | $< 10^{-11}$ | 0.74 |
| PMB | 1.08 (0.12) | 1.34 (0.16) | 208 | $< 10^{-10}$ | 0.69 |
| Global | 0.98 (0.09) | 1.23 (0.14) | 164 | $< 10^{-11}$ | 0.73 |
| Global pRNFL ~ whole disc pallor | 1.03 (0.11) | 1.26 (0.15) | 217 | $< 10^{-10}$ | 0.68 |
| Nasal-temporal ratio | 0.9 (0.07) | 0.9 (0.05) | 1063 | 0.82 | 0.02 |

**Abbreviations:** SD (standard deviation of the mean); PMB (Papillomacular bundle); pRNFL (peripapillary retinal nerve fibre layer).

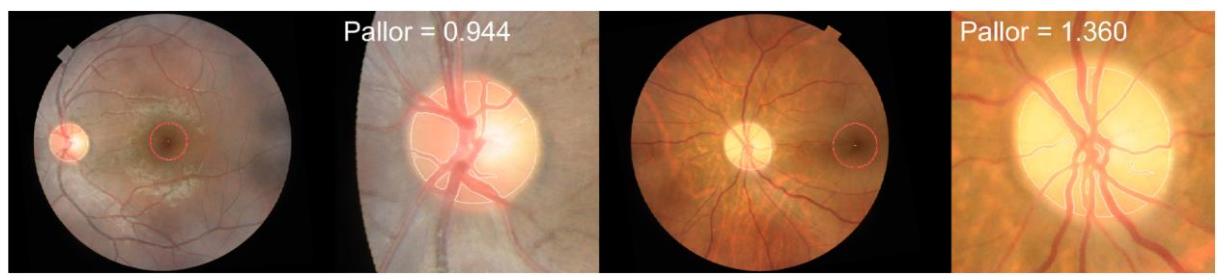

**Figure 17.** Intermittent stages of the pallor software on two images from the RFMiD dataset, one diagnosed by two ophthalmologists as having "optic disc pallor" (right) and a healthy control (left).

*Robustness to camera system, format, and resolution*

We tested the software on 5 different datasets (none of which were used in model development) containing images captured with 5 different camera systems (from three manufacturers), 3 different image formats, and resolutions ranging from 1634 × 1623 to 3072 × 2048. Judged by visual inspection, as per criterion described in the methods section, the methodology successfully processed all 50 images from all 5 datasets. Technical characteristics of the images are summarized in Table 6, and one example from each dataset is presented in Figure 18. Computation time was highest for the MESSIDOR dataset (2.9 seconds per image) and lowest for the REFUGE Canon dataset (2.5 seconds per image).

**Table 6.** Technical characteristics of the datasets used to assess how the software deals with images captured with a range of different camera systems, resolutions, and formats.

| Dataset | Format | Camera | Resolution | FOV | Centering | Dilation | Computation time per image (seconds) |
|---|---|---|---|---|---|---|---|
| G1020 | JPG | Topcon TRC-NW8 | between 1944 × 2108 and 2426 × 3007 | 45° | mixed | yes | 2.8 |
| MESSIDOR | TIFF | Topcon TRC-NW6 | 2240 × 1488 | 45° | mixed | yes | 2.9 |
| PREVENT (follow-up images) | BMP | Canon CR-Dgi | 3072 × 2048 | 45° | posterior pole | no | 2.7 |

| REFUGE (Canon) | JPG | Canon CR-2 | 1634 × 1634 | -- | posterior pole | -- | 2.5 |
| REFUGE (Zeiss) | JPG | Zeiss Visucam 500 | 2124 × 2056 | -- | posterior pole | -- | 2.6 |

**Abbreviations:** FOV (field of view).

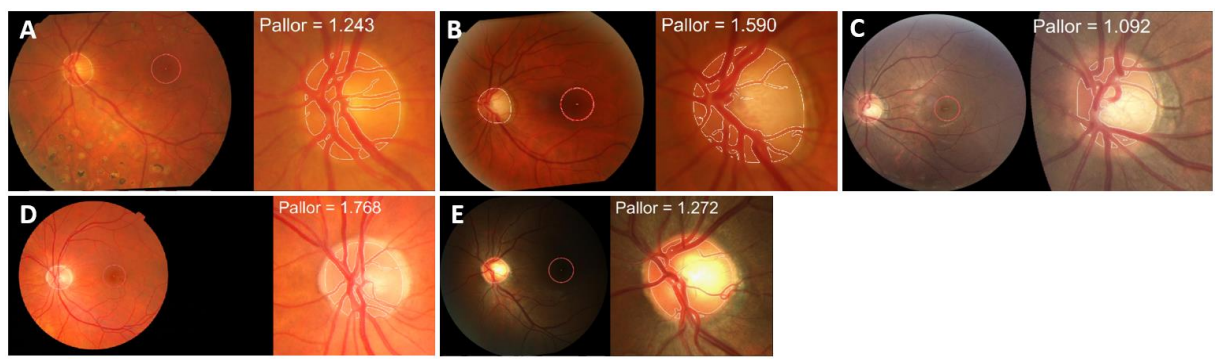

**Figure 18.** Robustness to camera system, format, and resolution results. A) G1020, B) MESSIDOR, C) REFUGE Canon, D) MESSIDOR, E) REFUGE Zeiss. Image to the left shows disc segmentation and fovea localization in the whole image, image to the right shows disc segmentation in closer detail. Pallor value shown is for the whole disc.

*Developing a set of automatic rejection criterion*

Of 1,584 images from the LBC dataset, 13 failed processing for reasons including excessive blur, optic disc outside field of view, and over/under-exposure. Rejection thresholds were set based on visual inspection of the remaining 1,571 images. Using our best judgement, thresholds for rejecting images were set at >.65 for eccentricity at > .65 AND < 50 for brightness of the control region. Examples of images that exceed these thresholds are presented in Figure 19. Summary statistics for images exceeding the proposed thresholds for both LBC and PREVENT are summarized in Table 7.

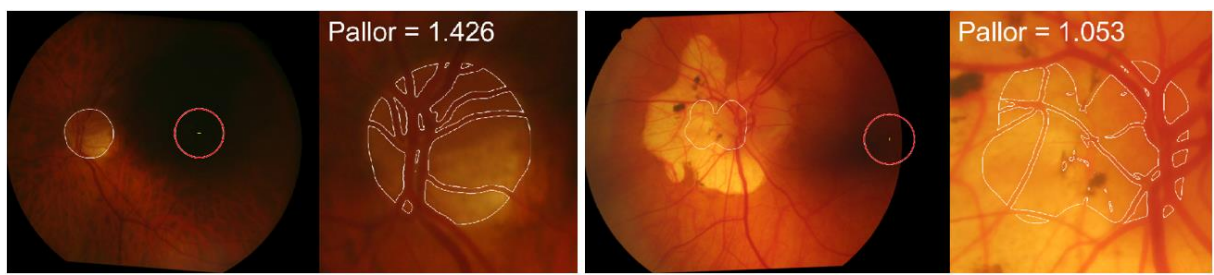

**Figure 19.** Automatic image rejection based on exceeding set thresholds for luminance (left) and eccentricity (right). We acknowledge that the image to the right was a failure of the software to correctly identify the disc margin due to excessive chorioretinal atrophy.

**Table 7.** Automatic image rejection thresholds.

| Criterion | Threshold | LBC | | PREVENT | |
|---|---|---|---|---|---|
| | | *Mean (SD)* | *Images over threshold (n, %)* | *Mean (SD)* | *Images over threshold (n, %)* |
| Eccentricity | > .65 | 0.39 (0.11) | 22 (1.4%) | 0.4 (0.1) | 2 (0.6%) |
| Luminance | < 50 | 110.21 (26.24) | 18 (1.15%) | 90.56 (18.62) | 9 (2.85%) |

**Abbreviations:** SD (standard deviation)

## Discussion

We have presented a fully automatic method of quantifying optic disc pallor in colour fundus photographs. In around 3 seconds per image, the software generates tabular data and visualizations capturing key measurements and summative properties. In particular, the software generates a global pallor metric, as well as metrics for 7 zones, in accordance with the Spectralis OCT peripapillary scan. The software proved robust to camera system, image format, and resolution in our experiments, and generates several metrics that can be used to filter out challenging or low-quality images, thereby allowing for application to large datasets.

In similar work, Yang *et al.*[31] developed a fully automatic pallor quantification system that operates on standard fundus photographs. However, their work has some limitations. For example, vasculature is included in their measurement region. This may be problematic, as vessel appearance is known to change with disease. For example, in hypertensive retinopathy, the arteriolar light reflex is accentuated,[41] in retinal vasculitis a white cuff is visible around vessels,[42] and although rare, in lipemia retinalis, vessels appear creamy.[43] In addition, zones in Yang's work (clock-hour locations) were not defined by their spatial relation to the fovea, making it difficult to accurately compare measurements across different images or to make sectoral comparisons to OCT. Our approach addresses these limitations by a) detecting vessels and excluding them from both the measurement and control region, and b) rotating the image along the optic disc-fovea axis prior to analysis.

In other similar work, Gonzalez-Hernandez et al.[32] developed a fully automated system to assess haemoglobin content in the optic disc (Laguna-ONhE; Optic Nerve Head Evaluation), which partly explains pallor. As with Yang's system, the Laguna software did not define the measurement zone in relation to the fovea. However, unlike Yang and the current study, Laguna does attempt to segment the optic cup. While this carries the advantage of capturing the entire NRR (where possible), it may fail when the cup is not visible, which is often the case in fundus photographs. Indeed, numerous studies show that segmenting the cup is difficult,[4,28,44,45] although recent work has been more successful,[46] and it is moreover difficult to establish ground truth given inter-observer variability in locating the extent of optic disc cupping. For this reason, we chose instead to define the measurement region in accordance with Yang et al. - at a fixed distance inward from the disc margin, sacrificing potential accuracy for robustness.

We investigated the relationship between pallor and pRNFL thickness in participants for whom concurrent data were available. Controlling for age, sex, disc area, control region brightness, and multiple comparisons, we found statistically significant associations between pallor and global pRNFL thickness, with a significant association also observed in the temporal-inferior and temporal-superior zones and in the temporal/nasal ratio. pRNFL thinning (as measured with OCT) is associated with several negative health outcomes including glaucoma,[47] increased cardiovascular risk,[48] Alzheimer's Disease and mild cognitive impairment,[49] future cognitive decline,[50] increased risk of dementia,[51] small vessel disease,[52] and stroke.[53] However, OCT is not yet widely available. Our approach generates measures of disc pallor that are associated with pRNFL thickness from simple colour fundus photographs, which are much more widely available, potentially enabling the detection, monitoring and progression of diseases that involve pRNFL loss with this imaging technology.

Aside from its association with pRNFL thickness, the ability to quantify pallor may have additional value in differentiating the aetiology of structural changes to the optic nerve head; for example, in differentiating glaucomatous and non-glaucomatous optic neuropathy. While pRNFL thinning is seen in both conditions, cupping rather than pallor is typical of glaucomatous optic neuropathy, and the presence of clinically apparent pallor often triggers investigations for non-glaucomatous causes, including potential MRI scanning of the anterior visual pathway.[6, 54]

In all zones, pallor was slightly lower in the right eye compared to the left eye, and this largely corresponded with pRNFL thickness measured in equivalent zones. This observation is in agreement with other studies that found the RNFL to be consistently thicker in the right eye.[55–58] Cameron *et al.*[59] discussed the importance of interocular symmetry in health and disease, pointing out that the emergence of *asymmetry* may alert the ophthalmologist that glaucoma should be considered. Further, they review several studies that attempt to create thresholds for when RNFL asymmetry may be clinically meaningful for glaucoma diagnosis and progression. This observation further suggests that our measure of symmetry (IoPV), may find utility in glaucoma detection and diagnosis.

Another important use case for the software could be the identification, monitoring and progression of compressive optic neuropathy (CON), whereby a compressive lesion anywhere along the optic nerve or anterior visual pathway (anterior to the lateral geniculate body) causes axons to die, resulting in optic atrophy/pallor.[60] The ability to quantify sectoral pallor may provide additional value. For example, compression to the optic chiasm can cause pallor in the temporal and nasal zones – a condition known as "band" or "bow-tie" atrophy.[61] Therefore, the pattern of optic disc pallor may further help localization of the lesion. Of particular relevance may be the detection of optic pathway gliomas (OPG), which predominantly affect children (mean age at presentation = 8.8 years).[62] Assessment of vision is crucial in diagnosis, however young children will often not complain of vision loss, and instead present at a later stage with a headache/pain.[63] Given that disc pallor is present in around 60% of cases,[62,63] it is feasible that OPG could be automatically detected through routine fundus imaging, which is not typically viewed by an ophthalmologist. Further research should investigate the association between optic disc pallor, as measured with the current software, and various types of CON.

Disc pallor is also an important measure of chemotherapy success in paediatric OPG,[64] with the authors suggesting that the degree of pallor could be important. Indeed, complementary work investigating the visual outcomes of childhood OPG treated with radiotherapy found that "severe" disc pallor (compared to "mild") at diagnosis or follow-up may be associated with a negative prognosis.[65] Further work could re-evaluate such existing studies, substituting clinical notes on pallor for the continuous measures generated by our software (dependent on the availability of fundus images). However, care should be taken if investigating *change* in pallor, as it rarely improves,[64] therefore the direction of change will almost always be one-way.

OCT is the gold standard for assessing RNFL loss. However, compared with fundus imaging it is costly, requires greater operator training, and is less prevalent. Furthermore, to avoid movement artefacts to which OCT is prone, such as those cause by ocular saccades, blinks, changes in head position, or respiratory movements,[66] patients must maintain a steady focus on a fixed point for several 10's of seconds.. Therefore, obtaining high-quality OCT can be particularly challenging in individuals who may struggle with prolonged focus and steadiness, such as children,[67] frail elderly, or those with movement disorders. Owing to the speed of acquisition, fundus imaging is more likely to be successful in these groups. Pallor derived from fundus photographs could provide an indicator that further examination is required, and could be good alternative in groups where OCT scanning is not feasible.

Strengths of the current study included the networks' ability to accurately segment the optic disc to the inner edge of the border tissue. The disc margin is marked differently depending on the imaging modality through which it is observed. In OCT, the margin is marked at Bruch's Membrane Opening (BMO),[20] while in clinical ophthalmoscopy and fundus photography, it is defined as the inner edge of the border tissue.[20] There are many deep learning-based disc segmentation algorithms (for a review see Hasan *et al.*),[29] however most are trained on one or more of four open-source image sets, namely IDRiD, RIMONE, DRISHTI-GS,[68] and DRIVE.[69] This may be problematic for our requirement, because on close inspection of ground truth segmentations in IDRiD and RIMONE, we observed a noticeable departure from what we perceived to be the clinically defined margin (Figure 20). We believe this could be the result of averaging multiple annotations from different individuals to arrive at a ground truth, in which there is considerable disagreement. Such disagreements may have arisen due to the annotators either marking the boundary clinically, or inferring BMO-based information, for example the bend of vessels at the rim. Although disagreement between multiple annotators provides an important measure of confidence that must be considered when assessing an automatic system, averaging multiple annotations can lead to label noise.

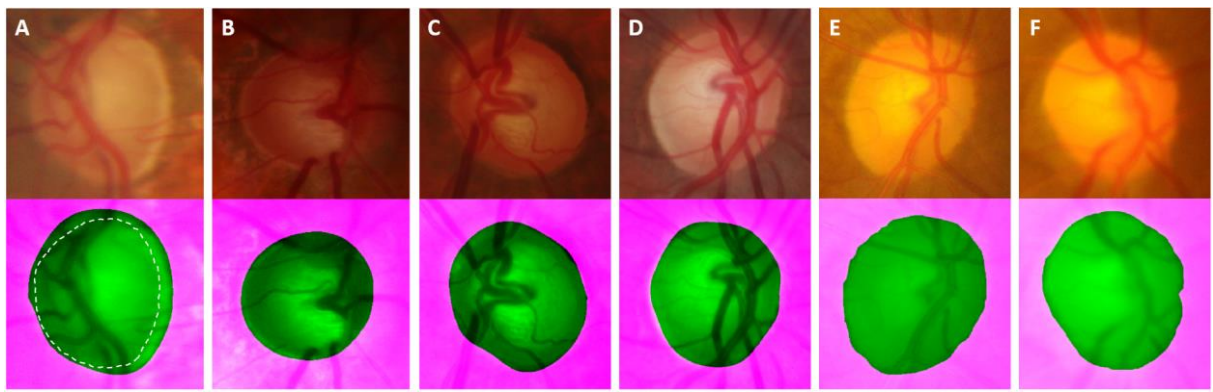

**Figure 20.** Examples from RIM ONE (glaucoma = A-B, non-glaucoma = C-D) and IDRiD (E-F) where the disc margin is over-estimated according to a clinical definition. The ground truth (according to the original annotations) is marked in pink. The dashed line in A represents where we perceive the margin to be. In A, the space between the dashed line and the start of the pink represents label noise.

A recent study on label noise in medical image segmentation demonstrated that while state-of-the-art networks are somewhat robust to unbiased, or random noise, they are sensitive to biased noise.[70] Indeed, we observed that RIMONE and IDRiD may contain biased noise, whereby the clinically defined disc boundary is systematically *over*-estimated with respect to where we perceive the clinical margin to lie. It is therefore possible that deep learning-based models trained on these datasets will systematically *over*-estimate the clinically defined boundary. The significance of this for automatic segmentation programs would depend on the application. For example, the VAMPIRE software,[71] which is concerned with obtaining vessel-based measurements, requires that the disc be estimated as a best-fit ellipse, in which case the precise boundary is less important. However, our work required greater precision, because sharp changes in colour at the border would erroneously affect pallor metrics. Another use case that may benefit from a more precisely defined disc margin is measuring the cup to disc ratio/profile in glaucoma,[72] because an over-estimated rim could erroneously widen the profile, leading to a false positive (missing glaucoma).

Another strength of the current study is the robustness of the fovea detection network, which, in our experiments, gives good estimates even for very challenging images (Figure 9). Additionally, locating the fovea helps the software determine which eye (left or right) is being processed, and allows for accurate zone placement. Lastly, the study had the advantage of using mixed effects modelling, which enabled the use of data from both eyes, thereby increasing statistical power.

The current study has several limitations. Firstly, disc appearance is affected by physiological factors, chiefly the media opacity of the lens, but also potentially the richness of the capillary net supplying the optic nerve. Lens status was not available for the patients included in our study, therefore we could not distinguish pallor caused by optic atrophy from pseudo pallor (non-pathological paleness, most notably caused by cataract extraction).[2] Further validation work should be carried out to assess the extent to which the current pallor metrics are affected by worsening cataract and cataract removal. In the meantime, information on cataract status, and other potential causes of pseudo pallor should be included as covariates where possible, particularly with older individuals. With regard to perfusion, in future studies, it would be interesting to examine the relationship between OCT-Angiography measures of vessel density and pallor measured using the software, to determine whether density of the capillary net might be a confounding factor in grading pallor, or if the software could potentially be useful for evaluating changes to the optic nerve head caused by reduced optic nerve head perfusion.

Another limitation is that pallor quantification was affected somewhat by the brightness of the control region (which itself is largely determined by pigmentation – a factor which varies among individuals. To overcome this, we added "control region brightness" as a covariate in all statistical models. While this may be an acceptable approach for research studies with multiple participants, for clinical insight into a single image, normative data with a wide range of relevant parameters would be required to determine the extent to which pigmentation affects the measure.

With regard to peripapillary atrophy, we selected the control region to include as much of the background retina as possible, while minimising the inclusion of any atrophy. Further, we used median values when calculating the overall control region brightness, which helped mitigate the effect of any included atrophy. However, if the control region of an image contains a significant portion of atrophy, the image should be rejected, as the pallor metrics may be unreliable (there were no such cases in the PREVENT dataset).

Another limitation is that 3 of the 4 networks (disc localisation and segmentation, and fovea localisation) were partly trained on PREVENT images. This calls into question the generalisability of the software. However, we point to internal (test set) and external testing carried out on all 3 networks, which show excellent generalisability. Nonetheless, further research should aim to replicate the finding that disc pallor is associated with pRNFL thickness in a novel dataset.

A final limitation was the lack of association between pRNFL thickness and pallor in 5 out of 7 zones. In similar work, but using clinical notes (e.g., pallor absent/present), Aleman *et al.*[5] observed that the association between pallor and pRNFL thickness is optimal only when significant thinning has occurred. This may help explain the lack of associations in 5 zones in the current study, as participants in the PREVENT cohort are relatively healthy; global mean pRNFL was thick (98.2μm, SD = 8.3) in comparison with normative data (90μm in whites;[73] 94μm in a multi-ethnic cohort),[73] although thinner than in individuals from Ghana (102μm).[74]

Further research should aim to replicate the current findings in a larger sample, generate normative data, and test for associations with cardiovascular risk factors and disease.

## Conclusions

A pale disc indicates irreversible damage to the anterior visual pathway and is present in numerous diseases. We present an automatic, AI-enabled method that is fast, easy to use, robust, and suitable for application to large datasets. We found associations between pallor and pRNFL thickness, suggesting that disc pallor derived from fundus photographs may act as a proxy for pRNFL. We think our method will be useful for the identification, monitoring and progression of diseases characterized by disc pallor/optic atrophy, including glaucoma, compression, and potentially in neurodegenerative disorders.

## Acknowledgements

We would like to thank all the PREVENT-Dementia participants for kindly donating their time.

# References

1. Ahmad SS, Kanukollu VM. Optic Atrophy. In: *StatPearls*. StatPearls Publishing; 2022. Accessed August 29, 2022. http://www.ncbi.nlm.nih.gov/books/NBK559130/

2. Osaguona VB. Differential diagnoses of the pale/white/atrophic disc. *Community Eye Health*. 2016;29(96):71-74.

3. Ahmad SS, Kanukollu VM. Optic Atrophy. *Handb Pediatr Retin OCT Eye-Brain Connect*. Published online May 5, 2022:292-295. doi:10.1016/B978-0-323-60984-5.00064-0

4. O'Neill EC, Danesh-Meyer HV, Kong GXY, et al. Optic disc evaluation in optic neuropathies: the optic disc assessment project. *Ophthalmology*. 2011;118(5):964-970. doi:10.1016/j.ophtha.2010.09.002

5. Aleman TS, Huang J, Garrity ST, et al. Relationship Between Optic Nerve Appearance and Retinal Nerve Fiber Layer Thickness as Explored with Spectral Domain Optical Coherence Tomography. *Transl Vis Sci Technol*. 2014;3(6):4. doi:10.1167/tvst.3.6.4

6. Ramm L, Schwab B, Stodtmeister R, et al. Assessment of Optic Nerve Head Pallor in Primary Open-Angle Glaucoma Patients and Healthy Subjects. *Curr Eye Res*. 2017;42(9):1313-1318. doi:10.1080/02713683.2017.1307415

7. Vilser W, Nagel E, Seifert BU, Riemer T, Weisensee J, Hammer M. Quantitative assessment of optic nerve head pallor. *Physiol Meas*. 2008;29(4):451-457. doi:10.1088/0967-3334/29/4/003

8. Assad A, Caprioli J. Digital image analysis of optic nerve head pallor as a diagnostic test for early glaucoma. *Graefes Arch Clin Exp Ophthalmol*. 1992;230(5):432-436. doi:10.1007/BF00175928

9. Nakano E, Hata M, Oishi A, et al. Quantitative comparison of disc rim color in optic nerve atrophy of compressive optic neuropathy and glaucomatous optic neuropathy. *Graefes Arch Clin Exp Ophthalmol*. 2016;254(8):1609-1616. doi:10.1007/s00417-016-3366-2

10. Kang S, Kim US amuel. Using ImageJ to evaluate optic disc pallor in traumatic optic neuropathy. *Korean J Ophthalmol KJO*. 2014;28(2):164-169. doi:10.3341/kjo.2014.28.2.164

11. Ritchie CW, Ritchie K. The PREVENT study: a prospective cohort study to identify mid-life biomarkers of late-onset Alzheimer's disease. *BMJ Open*. 2012;2(6):e001893. doi:10.1136/bmjopen-2012-001893

12. Ritchie CW, Ritchie K. The PREVENT study: a prospective cohort study to identify mid-life biomarkers of late-onset Alzheimer's disease. *BMJ Open*. 2012;2(6):e001893. doi:10.1136/bmjopen-2012-001893

13. Taylor AM, Pattie A, Deary IJ. Cohort profile update: The Lothian birth cohorts of 1921 and 1936. *Int J Epidemiol*. 2018;47(4):1042-1060. doi:10.1093/ije/dyy022

14. Zhang Z, Yin FS, Liu J, et al. ORIGA(-light): an online retinal fundus image database for glaucoma analysis and research. *Annu Int Conf IEEE Eng Med Biol Soc IEEE Eng Med Biol Soc Annu Int Conf*. 2010;2010:3065-3068. doi:10.1109/IEMBS.2010.5626137

15. Jin K, Huang X, Zhou J, et al. FIVES: A Fundus Image Dataset for Artificial Intelligence based Vessel Segmentation. *Sci Data*. 2022;9(1):475. doi:10.1038/s41597-022-01564-3

16. Fumero F, Alayon S, Sanchez JL, Sigut J, Gonzalez-Hernandez M. RIM-ONE: An open retinal image database for optic nerve evaluation. In: *2011 24th International Symposium on Computer-Based Medical Systems (CBMS)*. ; 2011:1-6. doi:10.1109/CBMS.2011.5999143

17. Porwal P, Pachade S, Kamble R, et al. Indian Diabetic Retinopathy Image Dataset (IDRiD): A Database for Diabetic Retinopathy Screening Research. *Data*. 2018;3(3):25. doi:10.3390/data3030025

18. Pachade S, Porwal P, Thulkar D, et al. Retinal fundus multi-disease image dataset (Rfmid): A dataset for multi-disease detection research. *Data*. 2021;6(2):1-14. doi:10.3390/data6020014

19. Strouthidis N, Yang H, Reynaud J, et al. Comparison of Clinical and Spectral Domain Optical Coherence Tomography Optic Disc Margin Anatomy. *Invest Ophthalmol Vis Sci*. 2009;50(10):4709-4718. doi:10.1167/iovs.09-3586

20. Chauhan BC, Burgoyne CF. From clinical examination of the optic disc to clinical assessment of the optic nerve head: a paradigm change. *Am J Ophthalmol*. 2013;156(2):218-227.e2. doi:10.1016/j.ajo.2013.04.016

21. YANG F, ZAMZMI G, ANGARA S, et al. Assessing Inter-Annotator Agreement for Medical Image Segmentation. *IEEE Access Pract Innov Open Solut*. 2023;11:21300-21312. doi:10.1109/access.2023.3249759

22. K Z. Contrast Limited Adaptive Histogram Equalization. *Graph Gems*. 1994;0:474-485.

23. Minaee S, Boykov Y, Porikli F, Plaza A, Kehtarnavaz N, Terzopoulos D. Image Segmentation Using Deep Learning: A Survey. *IEEE Trans Pattern Anal Mach Intell*. 2022;44(7):3523-3542. doi:10.1109/TPAMI.2021.3059968

24. Chen LC, Zhu Y, Papandreou G, Schroff F, Adam H. Encoder-Decoder with Atrous Separable Convolution for Semantic Image Segmentation. Published online August 22, 2018. doi:10.48550/arXiv.1802.02611

25. Sandler M, Howard A, Zhu M, Zhmoginov A, Chen LC. MobileNetV2: Inverted Residuals and Linear Bottlenecks. Published online March 21, 2019. doi:10.48550/arXiv.1801.04381

26. Chollet F. Xception: Deep Learning with Depthwise Separable Convolutions. Published online April 4, 2017. doi:10.48550/arXiv.1610.02357

27. Deng J, Dong W, Socher R, Li LJ, Li K, Fei-Fei L. ImageNet: A large-scale hierarchical image database. In: *2009 IEEE Conference on Computer Vision and Pattern Recognition*. ; 2009:248-255. doi:10.1109/CVPR.2009.5206848

28. Yu S, Xiao D, Frost S, Kanagasingam Y. Robust optic disc and cup segmentation with deep learning for glaucoma detection. *Comput Med Imaging Graph*. 2019;74:61-71. doi:10.1016/j.compmedimag.2019.02.005

29. Hasan MdK, Alam MdA, Elahi MdTE, Roy S, Martí R. DRNet: Segmentation and localization of optic disc and Fovea from diabetic retinopathy image. *Artif Intell Med*. 2021;111:102001. doi:10.1016/j.artmed.2020.102001

30. Al-Bander B, Al-Nuaimy W, Williams BM, Zheng Y. Multiscale sequential convolutional neural networks for simultaneous detection of fovea and optic disc. *Biomed Signal Process Control*. 2018;40:91-101. doi:10.1016/j.bspc.2017.09.008

31. Yang HK, Oh JE, Han SB, Kim KG, Hwang JM. Automatic computer-aided analysis of optic disc pallor in fundus photographs. *Acta Ophthalmol (Copenh)*. 2019;97(4):e519-e525. doi:10.1111/aos.13970

32. Gonzalez-Hernandez M, Gonzalez-Hernandez D, Perez-Barbudo D, Rodriguez-Esteve P, Betancor-Caro N, de la Rosa MG. Fully automated colorimetric analysis of the optic nerve aided by deep learning and its association with perimetry and oct for the study of glaucoma. *J Clin Med*. 2021;10(15). doi:10.3390/jcm10153231

33. MacGillivray TJ, Cameron JR, Zhang Q, et al. Suitability of UK Biobank Retinal Images for Automatic Analysis of Morphometric Properties of the Vasculature. *PLOS ONE*. 2015;10(5):e0127914. doi:10.1371/JOURNAL.PONE.0127914

34. Ying G shuang, Maguire MG, Glynn R, Rosner B. Tutorial on Biostatistics: Linear Regression Analysis of Continuous Correlated Eye Data. *Ophthalmic Epidemiol*. 2017;24(2):130-140. doi:10.1080/09286586.2016.1259636

35. VanderWeele TJ. Principles of confounder selection. *Eur J Epidemiol*. 2019;34(3):211-219. doi:10.1007/s10654-019-00494-6

36. Berendschot TTJM, DeLint PJ, Norren D van. Fundus reflectance—historical and present ideas. *Prog Retin Eye Res*. 2003;22(2):171-200. doi:10.1016/S1350-9462(02)00060-5

37. Bajwa MN, Singh GAP, Neumeier W, Malik MI, Dengel A, Ahmed S. G1020: A Benchmark Retinal Fundus Image Dataset for Computer-Aided Glaucoma Detection. Published online May 28, 2020. doi:10.48550/arXiv.2006.09158

38. Decencière E, Zhang X, Cazuguel G, et al. FEEDBACK ON A PUBLICLY DISTRIBUTED IMAGE DATABASE: THE MESSIDOR DATABASE. *Image Anal Stereol*. 2014;33(3):231-234. doi:10.5566/ias.1155

39. Orlando JI, Fu H, Breda JB, et al. REFUGE Challenge: A Unified Framework for Evaluating Automated Methods for Glaucoma Assessment from Fundus Photographs. *Med Image Anal*. 2020;59:101570. doi:10.1016/j.media.2019.101570

40. Camejo L, Noecker RJ. CHAPTER 14 - Optic nerve imaging. In: Stamper RL, Lieberman MF, Drake MV, eds. *Becker-Shaffer's Diagnosis and Therapy of the Glaucomas (Eighth Edition)*. Mosby; 2009:171-187. doi:10.1016/B978-0-323-02394-8.00014-0

41. Wong TY, Mitchell P. Hypertensive Retinopathy. *N Engl J Med*. 2004;351(22):2310-2317. doi:10.1056/NEJMra032865

42. Abu El-Asrar AM, Herbort CP, Tabbara KF. Differential Diagnosis of Retinal Vasculitis. *Middle East Afr J Ophthalmol*. 2009;16(4):202-218. doi:10.4103/0974-9233.58423

43. Kumar J, Wierzbicki AS. Lipemia Retinalis. *N Engl J Med*. 2005;353(8):823-823. doi:10.1056/NEJMicm040437

44. Fu H, Cheng J, Xu Y, Wong DWK, Liu J, Cao X. Joint Optic Disc and Cup Segmentation Based on Multi-Label Deep Network and Polar Transformation. *IEEE Trans Med Imaging*. 2018;37(7):1597-1605. doi:10.1109/TMI.2018.2791488

45. Sevastopolsky A. Optic disc and cup segmentation methods for glaucoma detection with modification of U-Net convolutional neural network. *Pattern Recognit Image Anal*. 2017;27(3):618-624. doi:10.1134/S1054661817030269

46. Meng Y, Zhang H, Zhao Y, et al. Graph-Based Region and Boundary Aggregation for Biomedical Image Segmentation. *IEEE Trans Med Imaging*. 2022;41(3):690-701. doi:10.1109/TMI.2021.3123567

47. Leung CKS, Choi N, Weinreb RN, et al. Retinal Nerve Fiber Layer Imaging with Spectral-Domain Optical Coherence Tomography: Pattern of RNFL Defects in Glaucoma. *Ophthalmology*. 2010;117(12):2337-2344. doi:10.1016/j.ophtha.2010.04.002

48. Chen Y, Yuan Y, Zhang S, et al. Retinal nerve fiber layer thinning as a novel fingerprint for cardiovascular events: results from the prospective cohorts in UK and China. *BMC Med*. 2023;21(1):24. doi:10.1186/s12916-023-02728-7

49. Thomson KL, Yeo JM, Waddell B, Cameron JR, Pal S. A systematic review and meta-analysis of retinal nerve fiber layer change in dementia, using optical coherence tomography. *Alzheimers Dement Diagn Assess Dis Monit*. 2015;1(2):136-143. doi:10.1016/j.dadm.2015.03.001

50. Ko F, Muthy ZA, Gallacher J, et al. Association of Retinal Nerve Fiber Layer Thinning with Current and Future Cognitive Decline: A Study Using Optical Coherence Tomography. *JAMA Neurol*. 2018;75(10):1198-1205. doi:10.1001/jamaneurol.2018.1578

51. Mutlu U, Colijn JM, Ikram MA, et al. Association of Retinal Neurodegeneration on Optical Coherence Tomography With Dementia: A Population-Based Study. *JAMA Neurol*. 2018;75(10):1256-1263. doi:10.1001/jamaneurol.2018.1563

52. Biffi E, Turple Z, Chung J, Biffi A. Retinal biomarkers of Cerebral Small Vessel Disease: A systematic review. *PLOS ONE*.:24.

53. Wang D, Li Y, Wang C, et al. Localized Retinal Nerve Fiber Layer Defects and Stroke. *Stroke*. 2014;45(6):1651-1656. doi:10.1161/STROKEAHA.113.004629

54. Conn FL MD, New Haven. When Glaucomatous Damage Isn't Glaucoma. Accessed April 17, 2023. https://www.reviewofophthalmology.com/article/when-glaucomatous-damage-isnt-glaucoma

55. Hwang YH, Song M, Kim YY, Yeom DJ, Lee JH. Interocular symmetry of retinal nerve fibre layer thickness in healthy eyes: a spectral-domain optical coherence tomographic study. *Clin Exp Optom*. 2014;97(6):550-554. doi:10.1111/cxo.12218

56. Dalgliesh JD, Tariq YM, Burlutsky G, Mitchell P. Symmetry of Retinal Parameters Measured by Spectral-domain OCT in Normal Young Adults. *J Glaucoma*. 2015;24(1):20. doi:10.1097/IJG.0b013e318287ac2f

57. Yang M, Wang W, Xu Q, Tan S, Wei S. Interocular symmetry of the peripapillary choroidal thickness and retinal nerve fibre layer thickness in healthy adults with isometropia. *BMC Ophthalmol*. 2016;16(1):182. doi:10.1186/s12886-016-0361-7

58. Budenz DL. Symmetry between the right and left eyes of the normal retinal nerve fiber layer measured with optical coherence tomography (an AOS thesis). *Trans Am Ophthalmol Soc*. 2008;106:252-275.

59. Cameron JR, Megaw RD, Tatham AJ, et al. Lateral thinking – Interocular symmetry and asymmetry in neurovascular patterning, in health and disease. *Prog Retin Eye Res*. 2017;59:131-157. doi:10.1016/j.preteyeres.2017.04.003

60. Rodriguez-Beato FY, De Jesus O. Compressive Optic Neuropathy. In: *StatPearls*. StatPearls Publishing; 2023. Accessed April 17, 2023. http://www.ncbi.nlm.nih.gov/books/NBK560583/

61. Monteiro MLR. Optical coherence tomography analysis of axonal loss in band atrophy of the optic nerve. *Br J Ophthalmol*. 2004;88(7):896-899. doi:10.1136/bjo.2003.038489

62. Fried I, Tabori U, Tihan T, Reginald A, Bouffet E. Optic pathway gliomas: a review. *CNS Oncol*. 2013;2(2):143-159. doi:10.2217/cns.12.47

63. Huang M, Patel J, Patel BC. Optic Nerve Glioma. In: *StatPearls*. StatPearls Publishing; 2023. Accessed April 18, 2023. http://www.ncbi.nlm.nih.gov/books/NBK557878/

64. Fisher MJ, Loguidice M, Gutmann DH, et al. Visual outcomes in children with neurofibromatosis type 1–associated optic pathway glioma following chemotherapy: a multicenter retrospective analysis. *Neuro-Oncol*. 2012;14(6):790-797. doi:10.1093/neuonc/nos076

65. Campagna M, Opocher E, Viscardi E, et al. Optic pathway glioma: Long-term visual outcome in children without neurofibromatosis type-1. *Pediatr Blood Cancer*. 2010;55(6):1083-1088. doi:10.1002/pbc.22748

66. Chhablani J, Krishnan T, Sethi V, Kozak I. Artifacts in optical coherence tomography. *Saudi J Ophthalmol*. 2014;28(2):81-87. doi:10.1016/j.sjopt.2014.02.010

67. Lee H, Proudlock FA, Gottlob I. Pediatric Optical Coherence Tomography in Clinical Practice—Recent Progress. *Invest Ophthalmol Vis Sci*. 2016;57(9):OCT69-OCT79. doi:10.1167/iovs.15-18825

68. Sivaswamy J, Krishnadas SR, Datt Joshi G, Jain M, Syed Tabish AU. Drishti-GS: Retinal image dataset for optic nerve head(ONH) segmentation. In: *2014 IEEE 11th International Symposium on Biomedical Imaging (ISBI)*. ; 2014:53-56. doi:10.1109/ISBI.2014.6867807

69. Niemeijer M, Staal J, Ginneken B, Loog M, Abramoff M. DRIVE: digital retinal images for vessel extraction. *Methods Eval Segmentation Index Tech Dedic Retin Ophthalmol*. Published online 2004.

70. Vorontsov E, Kadoury S. Label Noise in Segmentation Networks: Mitigation Must Deal with Bias. In: Engelhardt S, Oksuz I, Zhu D, et al., eds. *Deep Generative Models, and Data Augmentation, Labelling, and Imperfections*. Lecture Notes in Computer Science. Springer International Publishing; 2021:251-258. doi:10.1007/978-3-030-88210-5_25

71. Mookiah MRK, Hogg S, MacGillivray T, Trucco E. On the quantitative effects of compression of retinal fundus images on morphometric vascular measurements in VAMPIRE. *Comput Methods Programs Biomed*. 2021;202:105969. doi:10.1016/j.cmpb.2021.105969

72. MacCormick IJC, Williams BM, Zheng Y, et al. Correction: Accurate, fast, data efficient and interpretable glaucoma diagnosis with automated spatial analysis of the whole cup to disc profile (PLoS ONE (2019) 14:1 (e0209409) DOI: 10.1371/journal.pone.0209409). *PLoS ONE*. 2019;14(4):1-20. doi:10.1371/journal.pone.0215056

73. Knight OJ, Girkin CA, Budenz DL, Durbin MK, Feuer WJ, Cirrus OCT Normative Database Study Group  for the. Effect of Race, Age, and Axial Length on Optic Nerve Head Parameters and Retinal Nerve Fiber Layer Thickness Measured by Cirrus HD-OCT. *Arch Ophthalmol*. 2012;130(3):312-318. doi:10.1001/archopthalmol.2011.1576

74. Ocansey S, Abu EK, Owusu-Ansah A, et al. Normative Values of Retinal Nerve Fibre Layer Thickness and Optic Nerve Head Parameters and Their Association with Visual Function in an African Population. *J Ophthalmol*. 2020;2020:e7150673. doi:10.1155/2020/7150673